\documentclass[pra,a4paper,10pt,twocolumn,aps,pra]{revtex4-1} 

\usepackage{graphicx}

\usepackage[T1]{fontenc} 
\usepackage{graphicx}
\usepackage{color}
\usepackage{colortbl}

\usepackage[dvipsnames]{xcolor}
\usepackage{xfrac}
\usepackage{amsmath} 

\usepackage{braket}
\usepackage{mathrsfs}
\usepackage{dsfont}
\usepackage{caption}
\usepackage{subcaption}

\usepackage{ragged2e}
\DeclareCaptionJustification{justified}{\justifying}
\newcommand{\be}{ \begin{equation} }
\newcommand{\ee}{ \end{equation} }

\newcommand{\bqa}{ \begin{eqnarray} }
\newcommand{\eqa}{ \end{eqnarray} }

\begin{document} 

\title{\boldmath High Fidelity Quantum Gates beyond spectral selection}


\author{K. C. Matthew Cheung}
\author{Florian Mintert}


\affiliation{Department of Physics, Imperial College London, London SW7 2AZ, United Kingdom}

\email{kwok.cheung13@imperial.ac.uk/kcmc2@cam.ac.uk}

\begin{abstract}
Driving a certain transition without inducing undesired transitions is an ubiquitous problem in quantum control and the implementation of quantum information processing.
This problem gets the more challenging the weaker the desired transition couples to the control field, and the denser the system's spectrum is.
With the explicit example of a trapped ion we show how temporally shaped driving helps to increase the fidelity of a gate operation beyond the regular spectral selection of resonantly driven transitions.
We chose the explicit example of side-band transitions, since those couple more weakly to a control field than carrier transitions.
Driving a sideband transition without carrier excitation thus allows us to test the limits of frequently employed control tools, and we discuss their potential and limitations.
\end{abstract}

\maketitle

\section{Introduction}\label{Introduction}
In the last two decades, quantum information science has become an active field of research due to its many potential applications in computation \cite{nielsen2002quantum,shor1994algorithms,deutsch1992rapid}, cryptography \cite{ekert1991quantum,bennett1984quantum,kak2006three} and precision measurement of fundamental constants \cite{fischer2004new,peik2005new}.
Trapped ions are one of the most promising systems to implement quantum algorithms \cite{cirac1995quantum,sorensen1999quantum,blatt2008entangled} and quantum simulations \cite{buluta2009quantum,blatt2012quantum,kim2010quantum}.
Trapped ions can be controlled with laser beams, and desired processes
can be selected by choosing the laser frequency in resonance with the desired transition.
Off-resonant transitions, however, can be neglected only with the bounds set by the energy-time uncertainty, and they can be come very relevant for strong driving and/or dense spectra.
In order to achieve high fidelity quantum gates
it is thus desirable to develop accurate manipulation schemes,
where undesirable transitions are strongly suppressed.


With the advancement in pulse shaping technique \cite{kawashima1995femtosecond,assion1998control,levis2001selective,judson1992teaching}, it is nowadays possible to generate laser pulse with desired temporal shape,
and optimal control theory \cite{peirce1988optimal,kosloff1989wavepacket,d2007introduction} can help us to identify those pulses that induce the desired dynamics in a system \cite{verdeny2014optimal,2018arXiv180507351, 2018arXiv180506806}.
Common optimal control methods such as the GRAPE algorithm \cite{khaneja2005optimal} and the Krotov algorithm \cite{krotov1993global} aim at driving systems toward desired properties at single instances in time, while {\it e.g.} quantum simulations call for implementing desired dynamics during a continuous time window.
When the system is under periodic driving,
its dynamics can be described by a time-independent effective Hamiltonian according to the Floquet theory \cite{floquet1883equations},
which facilitates the identification of optimal driving patterns.

Goal of this paper is to devise control schemes that result in dynamics of the resonantly driven level under suppression of all other transitions.
We will discuss this in the context of trapped ions because this problem occurs in those systems very naturally.
Most of the subsequent findings are, however, by no means specific to trapped ions and can also be transferred to any other quantum system with discrete level structure.

This paper is structured as follows. In section 2, we discuss and derive the theoretical tools needed to find the optimal pulse to simulate the red sideband transition. In section 3, we apply the theoretical results in section 2 and present the numerical simulations. Section 4 summarizes the main points important for future work on the topic.

\section{Theory}\label{Theory}
\subsection{Trapped Ion under Driving}\label{Trapped Ion Under Driving}
The internal (electronic) degree of freedom of a trapped ion can be modelled as a two-level system with the Hamiltonian $(\omega_{eg}/2)\sigma_z$ defined in terms of the Pauli matrix $\sigma_z=\ket{e}\bra{e}-\ket{g}\bra{g}$.
The motion of the ion is modelled in terms of a harmonic oscillator with Hamiltonian $\nu a^{\dagger}a$, where $a$ and $a^{\dagger}$ denote annihilation and creation operators respectively.

Transitions between electronic levels $\ket{g}$ and $\ket{e}$ can be driven in terms of electromagnetic fields.
Since photons carry momentum, the absorption of a photon from such a field implies a change not only of the electronic states, but also of the motional state.
The interaction between an electromagnetic field and an ion is described by \cite{haffner2008quantum}
\be
H_I(\Omega,\eta,\omega_{em})=\Omega \sigma_x \cos\big(\eta(a+a^{\dagger})-\omega_{em} t\big)
\ee
with the Pauli matrix $\sigma_x=\ket{e}\bra{g}+\ket{g}\bra{e}$, the Rabi-frequency $\Omega$, characteristic frequency $\omega_{em}$ of the electromagnetic field, and the Lamb-Dicke-parameter $\eta$ characterizing the strength of the coupling to the motional degree of freedom.

In the following, we will consider poly-chromatic driving, such that the system Hamiltonian reads
\be
H=\frac{\omega_{eg}}{2}\sigma_z+\nu a^{\dagger}a+\sum_jH_I(\Omega_j,\eta_j,\omega_j)\ 
\label{eq:Hi}
\ee
Since we aim at an analytic description, based on perturbative methods, it is essential to work in a suitable frame, that is defined in terms of the transformation
\be
U_{I}=\text{exp}\left(it\left[\frac{\omega_{eg}+\Delta}{2}\sigma_z+\nu a^{\dagger}a\right]\right)
\ee
where the frequency-offset $\Delta$ will allow us to compensate
frequency shifts, such as the Lamb shift.
The transformed Hamiltonian $\bar H=U_{I} HU_I^{\dagger}+i \dot U_I U_I^\dagger$
reads
\be \label{eq:full_hamiltonian}
\bar H=-\frac{\Delta}{2}\sigma_z+\sum_j\bar H_I(\Omega_j,\eta_j,\omega_j)\ 
\ee
with
\be \label{eq:full_hamiltonian_supp}
\bar H_I(\Omega_j,\eta_j,\omega_j)=\Omega_j \sigma_x(t)  \cos(\eta_j(a(t)+a^{\dagger}(t))-\omega_j t)\ 
\ee
defined in terms of the time-dependent operators
\be
\begin{split}
\sigma_x(t)&=\sigma_+e^{i(\omega_{eg}+\Delta)t}+\sigma_-e^{-i(\omega_{eg}+\Delta)t} \\
a(t)&=ae^{-i\nu t}\ \\
a^\dagger(t)&=a^\dagger e^{i\nu t}\ 
\end{split}
\ee

We consider that the system is cooled to the Lamb-Dicke regime with a low-lying motional quantum number $k$, where $\eta_j^2(2k+1)$ is well below unity. In the Lamb-Dicke limit (where $\eta$ is sufficiently small), Eq. \eqref{eq:full_hamiltonian_supp} can be decomposed into terms of different powers in $\eta$ and approximated by its lowest two orders
\be
\begin{split}
\bar H_I(\Omega_j,\eta_j,\omega_j)&\approx \Omega_j \sigma_x(t)\Big(\mathds{1}\cos(\omega_jt)\\
&\quad+\eta_j(a(t)+a^{\dagger}(t))\sin(\omega_j t)\Big)\ 
\label{eq:ld_hamiltonian}
\end{split}
\ee
The lowest order term describes the carrier transition, {\it i.e.} transition between the electronic levels with no changes in the motional state. First order terms in $\eta$ describe side-band transitions with creation or annihilation of a motional quantum for every transition between $\ket{g}$ and $\ket{e}$. Overall, under the Lamb-Dicke approximation, the system Hamiltonian simplifies to
\begin{equation} \label{eq:simplified_hamiltonian}
\bar{H}=-\frac{\Delta}{2} \sigma_{z} +\Big(h_1(t)\sigma_{+}+h_2(t)\sigma_{+}a+h_3(t)\sigma_{+}a^{\dagger}+h.c.\Big)
\end{equation}
Depending on the detail of the polychromatic driving scheme (see subsection \ref{Poly-Chromatic Driving Scheme}), the three time-dependent terms ($h_1(t)$, $h_2(t)$ and $h_3(t)$) can all be found as Fourier sums by expanding Eq. \eqref{eq:ld_hamiltonian}. The first term in the Hamiltonian (Eq. \ref{eq:simplified_hamiltonian}) induces a phase shift to the system due to the transformation to a new frame. The three other terms in the Hamiltonian ($\sigma_{+}$, $\sigma_{+}a$ and $\sigma_{+}a^{\dagger}$) represent the carrier transition, the red sideband transition and the blue sideband transition respectively. 

For sufficiently weak driving, Eq.~\eqref{eq:Hi} can be reduced to its energy-conserving terms, so that carrier or side-band transitions can be realised simply by choosing the driving frequency to be on resonance with the desired transition, while
all energy-violating terms are then neglected under both Lamb-Dicke and rotating wave approximations. The identification of driving that can induce fast, high-fidelity quantum gates, however requires to take into account also off-resonant terms. As a result, we aim to construct polychromatic pulses to optimally simulate the red sideband transition by suppressing those undesirbale transitions.

\subsection{Floquet-Magnus Expansion}\label{Floquet-Magnus}

%


The dynamics of the driven system induced by the Hamiltonian (Eq. \eqref{eq:simplified_hamiltonian}) can be solved in terms of the Magnus expansion, in which the propagator $\bar U(t)$ satisfying $i\partial_t \bar U=\bar H(t) \bar U(t)$ is constructed in terms of the series
\be
\bar U(t)=\exp\big(-iM(t)\big)\ .
\ee
where $M(t)=\sum_l M_l(t)$. Each term in this Magnus expansion can be constructed in terms of a $l$-fold integral. The lowest order terms read \cite{magnus1954exponential,blanes2010pedagogical}
\begin{equation}\label{eq:magnus_terms}
\begin{split}
{M}_0(t)&=\int_{0}^td\tau \bar{H}(\tau)\\
{M}_1(t)&=-\frac{i}{2}\int_{0}^td\tau [\bar{H}(\tau),{M}_0(\tau)]\\
{M}_2(t)&=-\frac{i}{3}\int_{0}^td\tau [\bar{H}(\tau),{M}_1(\tau)]\\
&-\frac{1}{6}\int_{t\ge t_1\ge t_2\ge t_3\ge 0}\hspace{-1.6cm}dt_{1}dt_2dt_3\hspace{0.3cm} [\bar{H}(t_3),[\bar{H}(t_2),\bar{H}(t_1)]]\ 
\end{split}
\end{equation}
Restriction to the lowest order $M_0$ amounts to the rotating wave approximation.
In addition to renormalizing amplitudes for processes existing in the system Hamiltonian, ${M}_1(t)$ (and higher order terms) contain also terms describing further processes which are not contained in the underlying Hamiltonian. For example, ${M}_1(t)$ contains terms that describe coupling of the motional degrees of freedom to the qubit in terms of operators of the form $\sigma_z(a+a^{\dagger})$,
$\sigma_za^{\dagger}a$ and
$\sigma_z(a^2+(a^\dagger)^2)$.

Furthermore, we assume that the system is driven periodically, such that $\bar{H}(t+T)=\bar{H}(t)$. Under the assumption of periodicity, the propagator $\bar U(t)$ admits a simplification via the Floquet decomposition $\bar{U}(t)=U_F(t)\text{exp}(-iH_{eff}t)$, where $U_F$ is a $T$-periodic unitary operator satisfying $U_F(0)=U_F(T)=\mathds{1}$ and $H_{eff}$ is referred as the effective Hamiltonian \cite{floquet1883equations}. Under weak driving, the dynamics generated by the effective Hamiltonian neglects local fluctuations, but gives a global description of the dynamics. Using the Floquet theorem, the effective Hamiltonian can be straightfowardly determined via $H_{eff}=M(T)/T$. In principle, the Hamiltonian of the system scales with the driving amplitude ($\{\text{f}_j\omega\}$), such that the $l^{\text{th}}$ order Magnus term $M_l$ scales exactly of order ($\{(\text{f}_j)^l\}$). By matching the order of approximation, the effective Hamiltonian is expanded perturbatively in the following way,
\begin{equation}
H_{eff}=\frac{M_1(T)+M_2(T)+\cdots}{T}=H_{eff}^{(0)}+H_{eff}^{(1)}+\cdots
\end{equation}
where each term is obtained via the relationship $H_{eff}^{(l)}=M_{l+1}(T)/T$, and the analytical expressions of the three lowest order terms of the expanded effective Hamiltonian are provided in Appendix \ref{appendix:Effective Hamiltonian}.


Readers might notice that the system Hamiltonian (Eq. \eqref{eq:simplified_hamiltonian}) is not necessarily periodic (depending on the choices of $\omega_j$, $\nu$ and $\omega_{eg}+\Delta$). In subsection \ref{Poly-Chromatic Driving Scheme}, we will explain our polychromatic-driving scheme in more detail, and make use of realistic approximations to ensure that the system Hamitlonian is $T$-periodic.

\subsection{Target Dynamics and Target Functional}\label{Target Functional}
Since the Rabi-frequency for sideband transitions is reduced by a factor of $\eta$ as compared to the Rabi-frequency for carrier transitions,
realising a sideband transition without undesired contributions of off-resonantly driven carrier transitions is substantially more difficult than the process with reversed roles.
We will therefore focus to the realisation of dynamics that is induced by Hamiltonians of the form
\begin{equation}\label{eq:target_hamitlonian}
H_{tg}=\frac{i\text{f}_{tg}\omega}{2}\left(\sigma_+a-\sigma_-a^{\dagger}\right)
\end{equation}
where $\text{f}_{tg}\omega$ is the Rabi-frequency of the targeted red sideband transition, and $\omega=2\pi/T$ is the fundamental frequency of a $T$-periodic system.

A target functional can be defined to measure the performance of a control pulse. By optimizing the target functional, we can determine an optimal pulse which drives the system toward desired properties, such as maximizing the population transfer of a target state \cite{kumar2011optimal,boscain2002optimal} or minimizing the effect of decoherence in an open quantum system \cite{haddadfarshi2016high,cui2008optimal}. The common way of defining a target functional is based on the concept of either gate infidelity or state infidelity. In the following, we will define both gate infidelity and state infidelity to compare which definition is more suitable in simulating the target dynamics (Eq. \eqref{eq:target_hamitlonian}). The gate infidelity is defined as \cite{verdeny2014optimal},
\begin{equation}\label{eq:gate_infidelity}
\begin{split}
\mathcal{I}_{gate}=\frac{1}{T}\int^{T}_0 dt \, ||\bar{U}-U_{tg}||^2
\end{split}
\end{equation}
where $||A||^2=\mbox{tr}(A^{\dagger}A)$ is known as the Hilbert-Schmidt norm, and $\mbox{tr}( )$ denotes the trace of an operator. The gate infidelity (Eq. \eqref{eq:gate_infidelity}) is approximated to the lowest non-trivial order, and the driving amplitudes of the poly-chromatic pulse is constrained to the same order of approximation such that 
\begin{equation}\label{eq:eff_constraint}
H_{eff}=H_{tg}
\end{equation}
This results in a simplification of the above expression for the gate infidelity,
\begin{equation}\label{eq:gate_infidelity_simplified}
\begin{split}
\mathcal{I}_{gate}&\approx\frac{1}{T}\int^{T}_0 dt \, \mbox{tr}(2\mathds{1}-U_F-U_F^{\dagger})\\
&\approx\frac{1}{T}\int^{T}_0 dt \, \mbox{tr}\Big(\big(M_1(t)-H_{eff}^{(0)}t\big)^2\Big)+O\Big((\text{f}_j)^3\Big)\\
&=\frac{1}{T}\int^{T}_0 dt \, \sum_{k=0}^{\infty}\Big(2|g_1|^2+(2k+1)(|g_2|^2+|g_3|^2)\Big)
\end{split}
\end{equation}
where the analytical expressions of the integrals of $|g_1|^2$, $|g_2|^2$ and $|g_3|^2$ are provided in Appendix \ref{appendix:Target Functional}, and in terms of the physical meaning, the time-averaged integrals of $|g_1|^2$, $|g_2|^2$ and $|g_3|^2$ measure the deviation of the driven dynamics from the target dynamics in the carrier process, the red sideband process and the blue sideband process respectively. 

The dimension of a quantum harmonic oscillator is infinite, and therefore the above gate infidelity (Eq. \eqref{eq:gate_infidelity_simplified}) does not return a finite value. As a consequence, we choose to truncate the system to a finite-dimensional subspace spanned by $\{\ket{g,0},\ket{e,0},\ldots,\ket{g,d-1},\ket{e,d-1}\}$ and divide the functional by $\sum_{k=0}^{d-1} (2k+1)$, we then arrive the following asymptotic expression by taking the limit $d\rightarrow\infty$,
\begin{equation}\label{eq:gate_infidelity_asymptotic}
\mathcal{I}_{gate,asy}=\frac{1}{T}\int^{T}_0 dt \, \Big(|g_2|^2+|g_3|^2\Big)
\end{equation}
Since the Rabi-frequency of the carrier transition does not depend on the motional state, the time-averaged fluctuations in the carrier transition do not scale with the motional quantum number. In particular, the red and blue sideband dynamics fluctuate more rapidly than the carrier dynamics for high-lying motional states, and as a result, the asymptotic expression of the gate infidelity does not contain the time-averaged integral of $|g_1|^2$. However, trapped ions are often cooled to extremely low temperature and low-lying motional states are mostly occupied. At low temperature, the fluctuations in the carrier transition ($|g_1|^2$) are dominant, since both $|g_2|^2$ and $|g_3|^2$ are smaller by a scale factor of $\eta$. The asymptotic expression (Eq. \eqref{eq:gate_infidelity_asymptotic}) neglects the effect of fluctuations in the carrier transition, and this might result in a poor control performance (To be discussed further in subsection \ref{State Infidelity vs. Gate Infidelity}). 

An alternative would thus be to truncate to a finite dimensional subspace, consistently with the achievable cooling.
Having this in mind, we can first define target functionals
\begin{equation}\label{eq:state_infidelity}
\begin{split}
\mathcal{I}_{state}=\frac{1}{T}\int^{T}_0 dt \, \Big(1- |\braket{i|\bar{U}^{\dagger}U_{tg}|i}|^2\Big)
\end{split}
\end{equation}
for the set of initial states $\ket{i}$.
Using Eq. \eqref{eq:eff_constraint} again, the state infidelity can be simplified,
\begin{equation}\label{eq:state_infidelity_simplified}
\begin{split}
\mathcal{I}_{state}&\approx\frac{1}{T}\int^{T}_0 dt \, \Big(1- |\braket{i|U_{tg}^{\dagger}U_FU_{tg}|i}|^2\Big)\\
&\approx\frac{1}{T}
\begin{cases}
\int^{T}_0 dt \, \Big(|g_1|^2+k|g_2|^2+k|g_3|^2\\
\,\,\quad+\text{cos}(\text{f}_{tg}\sqrt{k}\omega t)|g_3|^2\Big),\,\,  \text{for} \ket{i}=\ket{g,k} \\
\int^{T}_0 dt \, \Big(|g_1|^2+k|g_2|^2+k|g_3|^2\\
\,\,\quad-\text{cos}(\text{f}_{tg}\sqrt{k}\omega t)|g_3|^2\Big), \,\, \text{for} \ket{i}=\ket{e,k-1}
\end{cases}
\end{split}
\end{equation}
The state infidelity has a similar structure to the gate infidelity apart from the interference factor ($\text{cos}(\text{f}_{tg}\sqrt{k}\omega t)$), and, most importantly, the state infidelity always returns finite values.
In the following, we will minimize the state infidelity (Eq. \eqref{eq:state_infidelity_simplified}) to determine the optimal pulse in section \ref{Numerical Results}, and only in subsection \ref{State Infidelity vs. Gate Infidelity}, we will compare the performance of optimal pulses found by minimizing the gate infidelity and the state infidelity respectively.


%
%
%
%
%

\subsection{Poly-Chromatic Driving Scheme}\label{Poly-Chromatic Driving Scheme}

We have so far outlined the general theory describing the interaction between trapped ions and laser fields. However, we have not presented the driving scheme, the explicit form of the system Hamiltonian (Eq. \eqref{eq:simplified_hamiltonian}) up to this point remains unspecified. In this subsection, we will explain the poly-chromatic driving scheme in more detail.

Typical manipulations are based on driving the system resonantly, however, in the Lamb-Dicke regime, resonant manipulation is often not ideal due to the Lamb shift. A solution to such problem is to drive the system off-resonantly by detuning the laser frequency from resonance slightly, such that the Lamb shift can be cancelled by the off-resonant term while the system is  driven on resonance effectively.


Following the above discussion, we propose a polychromatic driving scheme, where the frequency spectrum is not centered around the red sideband transition frequency ($\omega_{eg}-\nu$) but instead slightly detuned (to be centered around $\omega_{eg}-\nu+\Delta$). In order to apply the Floquet-Magnus expansion, we also need to make sure that the system is periodically driven. First of all, we require that $\nu$ is an integer multiple of the fundamental frequency $\omega$. With $\nu=m\omega$ (where $m$ is a positive integer), it eases the notation to write both $\Omega_j=\text{f}_j\omega$ and $\Delta=\delta \omega$ in terms of the fundamental frequency $\omega$, where both $\text{f}_j$ and $\delta$ are assumed to be real numbers. Under the driving scheme described above, $\omega_j$ is written as
\be
\omega_j = \omega_{eg}-m\omega +\delta \omega +j \omega
\ee
where $j$ is an integer running from $-n$ to $n$. Secondly, we expand Eq. \eqref{eq:ld_hamiltonian} and write the system Hamiltonian as a Fourier sum of all the frequencies in the poly-chromatic pulse. However, the Hamiltonian is still comprised of highly oscillatory terms such as $e^{i(2\omega_{eg}+(2\delta-m+j)\omega)t}$. Within an ion trap, the transition frequency of the qubit is often chosen to be much larger than the trap's motional frequency ($\omega_{eg}\gg m\omega$). As a result, the system Hamiltonian can further be simplified by applying the rotating wave approximation to drop all oscillatory terms with the factor $e^{2i\omega_{eg}t}$. Finally, the three time-dependent terms in Eq. \eqref{eq:simplified_hamiltonian} read
\begin{equation}\label{eq:time_dependent_terms}
\begin{split}
h_1(t)&=\sum_{j=-n}^{n}\frac{\text{f}_j \omega}{2}e^{i(m-j)\omega t}\\
h_2(t)&=\sum_{j=-n}^{n}\frac{i\eta_j\text{f}_j \omega}{2} e^{-ij\omega t}\\
h_3(t)&=\sum_{j=-n}^{n}\frac{i\eta_j\text{f}_j \omega}{2}  e^{i(2m-j)\omega t} 
\end{split}
\end{equation}
The overall idea is to find the set of optimal coefficients $\{\delta, \text{f}_j\}$, such that the driven system simulates the ideal red sideband dynamics (Eq.~\eqref{eq:target_hamitlonian}).

\section{Poly-Chromatically Driven Dynamics}\label{Numerical Results}
\subsection{Numerical Optimization of the Target Functional}\label{Numerical Optimization Of The Target Functional}
In this subsection, we will explain the minimization scheme in more detail before presentng the simulation results. To determine the set of optimal coefficients $\{\delta,\text{f}_j\}$, we use the sequential least squares programming algorithm \cite{kraft1988software} available in the Scipy packagae (http://www.scipy.org/) to minimize the state infidelity (Eq. \eqref{eq:state_infidelity_simplified}). The minimization of Eq. \eqref{eq:state_infidelity_simplified} is subject to constraints which are provided by the effective Hamiltonian calculation (See Appendix \ref{appendix:Effective Hamiltonian} for details). The effective Hamiltonian takes the following form,
\begin{equation}\label{eq: effective_ham}
\begin{split}
H_{eff}\approx&H_{eff}^{(0)}+H_{eff}^{(1)}+H_{eff}^{(2)}\\
=&c_{1}\mathds{1}+c_{2}\sigma_z+c_{3}\sigma_za+c_{4}\sigma_z a^2+c_{5}\sigma_za^{\dagger}a\\
&+c_{6}\sigma_++c_{7}\sigma_+a+c_{8}\sigma_+a^{\dagger}+h.c.\\&+O(\eta^2,\eta^3)
\end{split}
\end{equation}
where analytical expressions of above coefficients ($c_i$) are provided in Appendix \ref{appendix:Effective Hamiltonian}. As mentioned in subsection \ref{Target Functional}, the effective Hamiltonian is constrained to be equal to the target Hamiltonian. One intuitive choice of constraints would be $c_{7}=i\text{f}_{tg}\omega/2$, while setting all the other terms in Eq. \eqref{eq: effective_ham} to be equal to zero. One of the renormalized terms ($c_{1}\mathds{1}$) induces only a global phase shift to the system, and as a result, it is not necessary to constrain this term to be equal to zero. This gives seven constraints in total instead of eight, however, we have found that minimization result subject to the seven constraints leads to a smaller effective Rabi-frequency when compared to the Rabi-frequency of the target dynamics. Instead, we have constrained the two energy shift terms ($\sigma_z$ and $\sigma_za^{\dagger}a$) and the two motional transition terms ($\sigma_za$ and $\sigma_z a^2$) to be equal to zero, and we have also constrained the red sideband terms ($\sigma_+a$) to be equal to the target Hamiltonian, while the two other electronic transitions ($\sigma_+$ and $\sigma_+a^{\dagger}$) have not been constrained (Further details are provided in Eq. \eqref{eq:constraints} in Appendix \ref{appendix:Effective Hamiltonian}).

As a result, the state infidelity (Eq. \eqref{eq:state_infidelity_simplified}) is minimized subject to five constraints, and in subsection \ref{Work_Less_Optimal}, we will explain briefly why the above choice of five constraints works better than the more complete choice of seven constraints. Up to this point, the numerical value of the detuning ($\delta$) has not been specified. The minimization of the infidelity is subject to constraints which are dependent of $\delta$, such that different sets of optimal coefficients will be found by varying $\delta$. To determine the one set of optimal coefficients $\{\text{f}_j\}$, $\delta$ is varied till the target functional reaches the minimal value among different input values of $\delta$.

\subsection{Comparison with Monochromatic Driving}\label{Comparison With Monochromatic Driving}
\begin{figure*}
\centering
\begin{subfigure}[t]{0.5\textwidth}
\centering
\includegraphics[width=8.6cm]{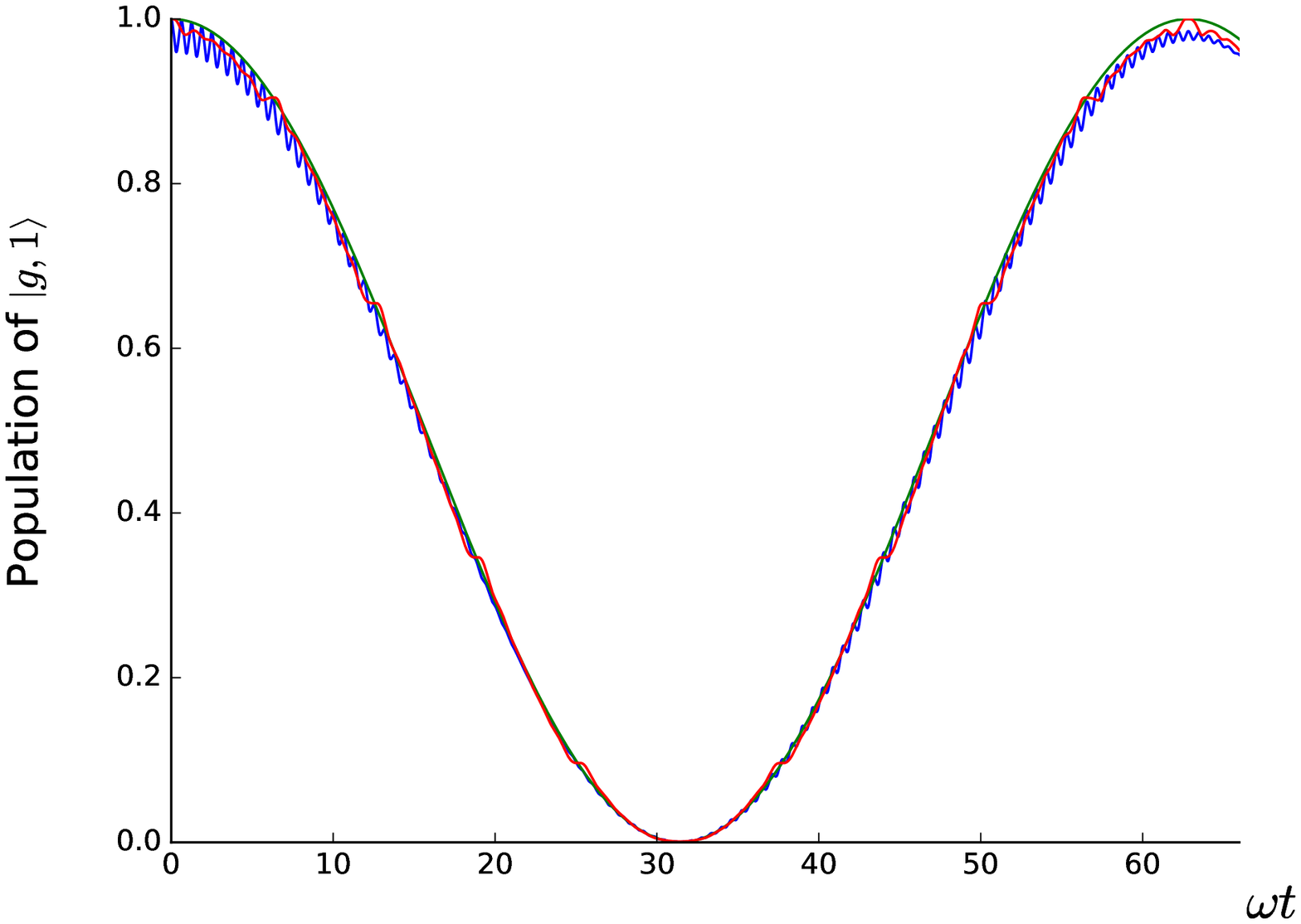}
\caption{\label{fig:k=1_g1}}
\end{subfigure}%
\hfill
\begin{subfigure}[t]{0.5\textwidth}
\centering
\includegraphics[width=8.6cm]{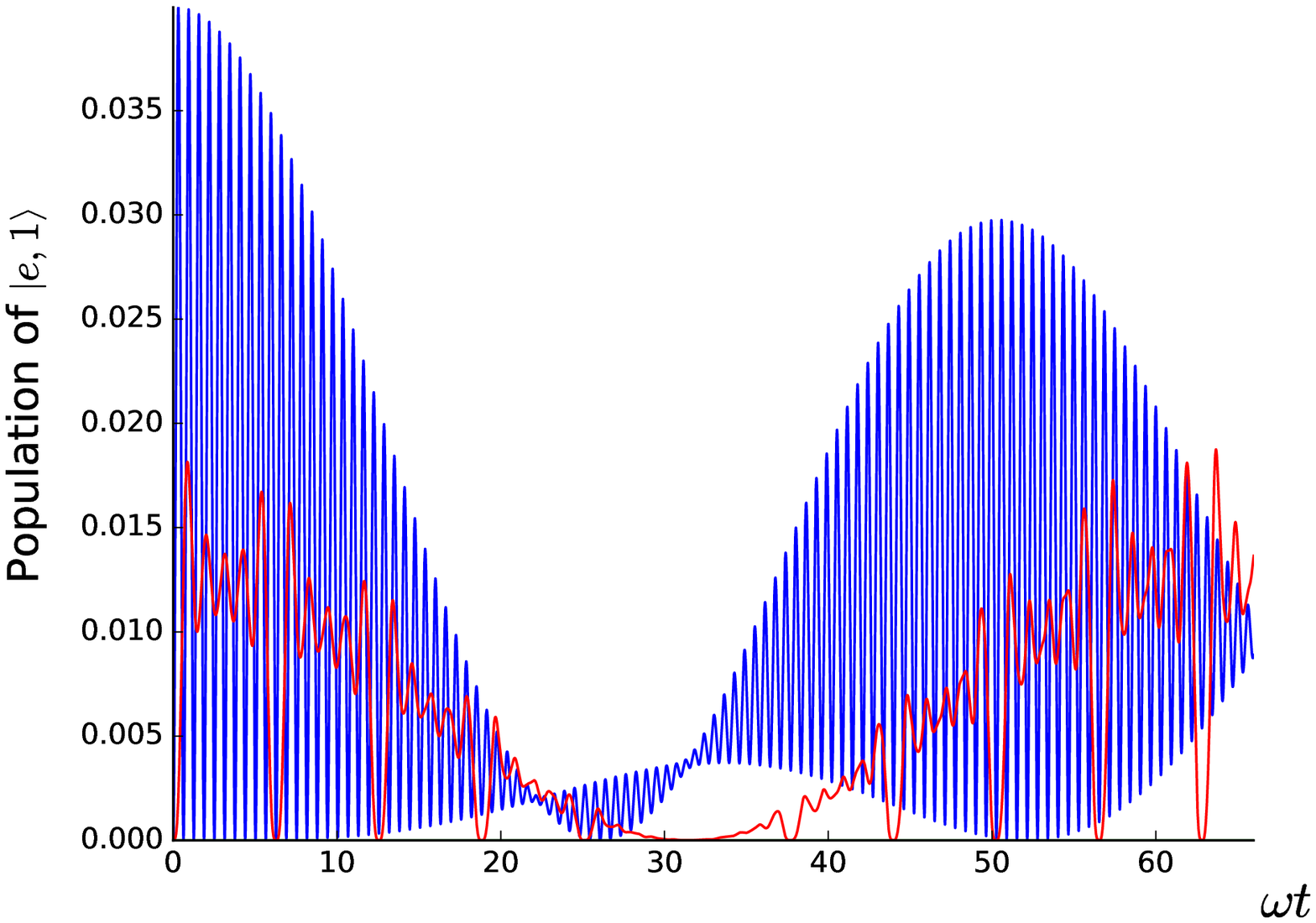}
\caption{\label{fig:k=1_e1}}
\end{subfigure}%
\caption{\label{fig:k=1_weak}Monochromatic driving vs. poly-chromatic driving. Populations (a) $P_{g1}(t)=\mbox{tr}(\ket{g,1}\bra{g,1}\rho(t))$ (b) $P_{e1}(t)=\mbox{tr}(\ket{e,1}\bra{e,1}\rho(t))$ for the monochromatic dynamics (blue line), the target dynamics (green line) and the poly-chromatic dynamics (red line). Parameters: $\text{f}_{tg}=0.1$, $m=10$, $\eta=0.05$ (Lamb-Dicke parameters are assumed to be the same for all frequencies) and $n=5$.}
\end{figure*}

In the Lamb-Dicke limit, the Lamb shift term ($\sigma_z$) is not negligible. To keep driving the system on resonance effectively, a monochromatic pulse must be detuned by $(\text{f}_{tg})^2/2\eta^2m$, such that the first order Lamb shift is cancelled. However, there are other high order terms in the effective Hamiltonian (Eq. \eqref{eq: effective_ham}) that can induce undesirable transitions. In this subsection, we will show how the optimal poly-chromatic pulse can improve the control of the sideband transition.

In Fig. \ref{fig:k=1_weak}, the system is initialised in an intial state $\rho_0=\ket{g,1}\bra{g,1}$. In Fig. \ref{fig:k=1_g1}, both monochromatically- and poly-chromatically-driven dynamics follow the target dynamics faithfully. For an ideal red sideband process with an initial state $\rho_0=\ket{g,1}\bra{g,1}$, the populations of $\ket{e,1}$ and $\ket{e,2}$ are expected to be zero. Under the driven Hamiltonian, both $\ket{e,1}$ and $\ket{e,2}$ can be excited via the carrier transition and the blue sideband transition respectively. As can be seen from Fig. \ref{fig:k=1_e1}, the carrier transition under poly-chromatic driving is clearly suppressed when comparing to the monochromatic case (the blue sideband transition is also suppressed, but is not shown here).

\begin{figure}
\centering
\includegraphics[width=8.6cm]{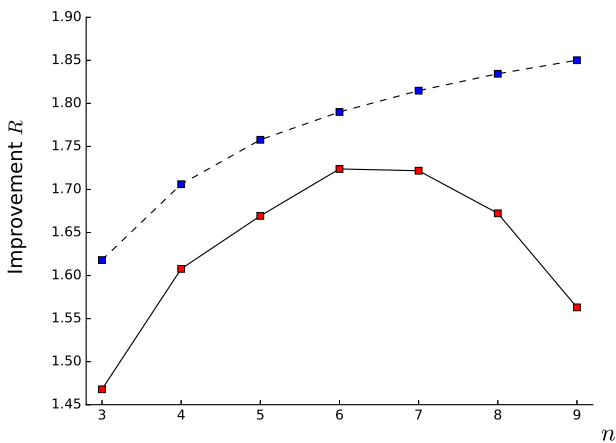}
\caption{\label{fig:improvement} Improvement $R$ plotted as a function of $n$. The dashed line (with blue square) represents the theoretical improvement corresponding to the minimization results of the perturbed functional (Eq. \eqref{eq:state_infidelity_simplified}), while the solid line (with red square) represents the actual improvement under one complete cycle of driving (Eq. \eqref{eq:infidelity_cycle}). Parameters: $\text{f}_{tg}=0.1$, $m=10$ and $\eta=0.05$ (Lamb-Dicke parameters are assumed to be the same for all frequencies).}
\end{figure}

In subsection \ref{Target Functional}, the state infidelity (Eq. \eqref{eq:state_infidelity}) is defined as an integral from $0$ to $T$, and is expanded perturbatively to arrive a simplified expression (Eq. \eqref{eq:state_infidelity_simplified}). The optimal coefficients $\{\delta, \text{f}_j\}$ are determined by minimiziming Eq. \eqref{eq:state_infidelity_simplified} as discussed in subsection \ref{Numerical Optimization Of The Target Functional}. Poly-chromatic driving is expected to perform better than monochromatic driving if the minimization of Eq. \eqref{eq:state_infidelity_simplified} is successful, however, Eq. \eqref{eq:state_infidelity_simplified} can only reflect the performance of the poly-chromatic pulse for a short interval of time (up to the point where second order approximation is valid). As we are interested in the control of a coherent quantum gate, it is reasonable to also measure the performance for a complete cycle ($\omega T_{cycle}=4\pi/\text{f}_{tg}$ for the sideband transition with quantum motional number $k=1$);
\begin{equation}\label{eq:infidelity_cycle}
\mathcal{I}_{{cylce}}=\frac{1}{T_{cycle}}\int^{T_{cycle}}_0 dt \, \Big(1- |\braket{i|\bar{U}^{\dagger}U_{tg}|i}|^2\Big)
\end{equation}
where the above quantity ($\mathcal{I}_{{cycle}}$) is determined numerically in the simulation to reflect the performance of the found optimal coefficients during one cycle of gate operation. Using the same definition in \cite{haddadfarshi2016high}, the improvement $R$ of the polychromtic pulse is defined to be the ratio of the gate infidelities (Eq. \eqref{eq:infidelity_cycle}) of the monochromatic pulse to the poly-chromatic pulse $\mathcal{I}_{{m}}/\mathcal{I}_{{p}}$. 

As can be seen from the dashed line depicted in Fig. \ref{fig:improvement}, the minimization results of Eq. \eqref{eq:state_infidelity_simplified} predict that the performance of the poly-chromatic pulse can be improved by increasing the number of frequencies. Meanwhile, the solid line depicted in Fig. \ref{fig:improvement} shows the exact performance of the optimal pulse under one complete cycle of driving, and the performance only gets improved by increasing from $n=3$ to $n=6$ and begins to deteriorate if one further increases the number of frequencies. By increasing $n$, the solution space has more degrees of freedom, such that the optimization is more flexible to reach a better optimal point when comparing with a smaller $n$, which is what the theoretical results has predicted. However, what Fig. \ref{fig:improvement} has shown is that the one-cycle simulations disagree with the theoretical predictions that the performance should improve with an increasing $n$. To explain the disagreement, we need to understand how the approximations made in section \ref{Theory} can affect the dynamics. In the evaluation of both second order and third order Magnus expansions, one needs to integrate exponential terms such as $e^{i(m-j)\omega t}$, and by increasing $n$ close to $m$, some terms like $e^{i(m-n)\omega t}$ are getting closer to resonance. As a result, the poly-chromatic pulse contains some slow frequency terms which are not present in the monochromatic pulse, and the expansion at second/third order becomes insufficient to approximate the actual dynamics. In Fig. \ref{fig:improvement}, the trap frequency ($m\omega$) is chosen to be equal to $10\omega$, and as can be seen from the solid line in the plot, the performance begins to deteriorate from $n=6$ to $n=9$ as a consequence of the breakdown of the current approximation.

Despite the slight decline in performance for higher $n$, the poly-chromatic pulse can in fact improve the performance of the gate by approximately a factor of 1.7 (when $n=6$) as shown in Fig. \ref{fig:improvement}, and the performance of the optimal pulse can in principle be further improved by going beyond the second order approximation in the target functional and the effective Hamiltonian.

\subsection{Timing Error}\label{Periodic Control}

\begin{figure}
%
%
\includegraphics[width=8.6cm]{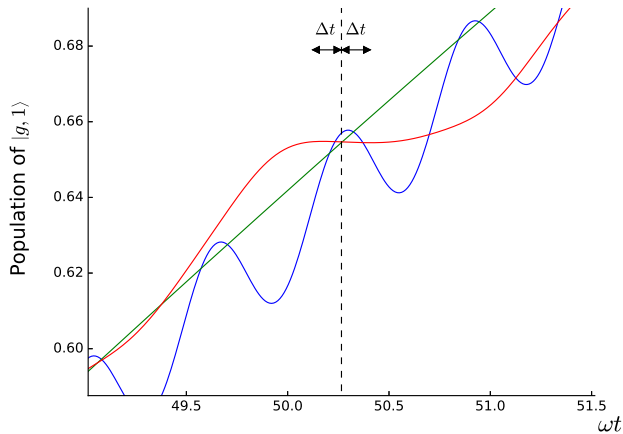}
\caption{\label{fig:k=1_zoomed}
Monochromatic driving vs. poly-chromatic driving over a shorter time window. Population $P_{g1}(t)=\mbox{tr}(\ket{g,1}\bra{g,1}\rho(t))$ during $7.8T\leq t \leq8.2T$ for the monochromatic dynamics (blue line), the target dynamics (green line) and the poly-chromatic dynamics (red line). The straight black dashed lines denote when $t$ is equal to the integral multiple of the period $T$, and $\Delta t$ denotes the random error of the switch-off time of the laser field. Parameters: $\text{f}_{tg}=0.1$, $m=10$, $\eta=0.05$ (Lamb-Dicke parameters are assumed to be the same for all frequencies) and $n=5$.}
\end{figure}

Under the constraint $H_{eff}=H_{tg}$ and the assumption of periodicity, the propagator of the system must be equal to the target propagator periodically $\bar{U}(qT)=U_{tg}(qT)$, for any integer $q$. As a result, the poly-chromatically driven dynamics is expected to be much closer to the target dynamics periodically than the monochromatically driven dynamics (in spite of the fact that some terms in the effective Hamiltonian are not included in the constraints).
This can be seen in Fig. \ref{fig:k=1_zoomed}, which depicts the population of $\ket{g,1}$ in the time-window around $t=8T$ (where $t=8T$ is an arbitrary choice), with the initial condition $\rho_0=\ket{g,1}\bra{g,1}$.
In addition to the general trend that the amplitude of fast oscillations in the poly-chromatically driven dynamics (red) is substantially lower than in the monochromatically driven dynamics (blue), one can see that the population of $\ket{g,1}$ is nearly constant in the time-window around $t=8T$. This implies that timing errors, {\it i.e.} fluctuations in switching the driving fields on and off when reaching a particular state at $t=qT$ (where $q$ is an integer), have a substantially lower detrimental impact in the case of optimised driving than in the case of regular monochromatic driving.

\subsection{Full Effective Hamiltonian Constraints Work Less Optimal}\label{Work_Less_Optimal}
\begin{figure*}
\begin{subfigure}[t]{0.5\textwidth}
\includegraphics[width=8.6cm]{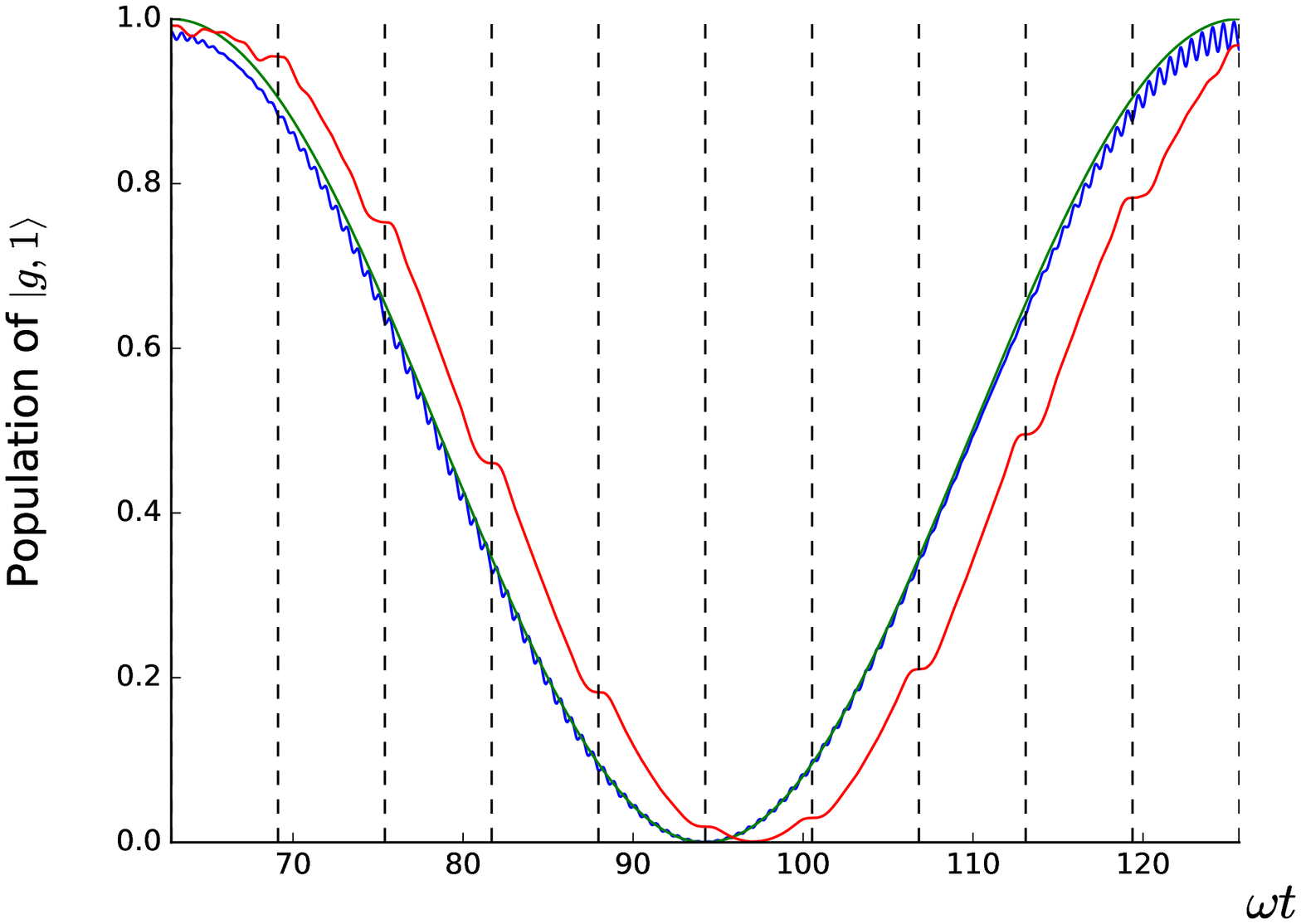} 
\caption{\label{fig:g1_full_constraints}}
\end{subfigure}%
\hfill
\begin{subfigure}[t]{0.5\textwidth}
\includegraphics[width=8.6cm]{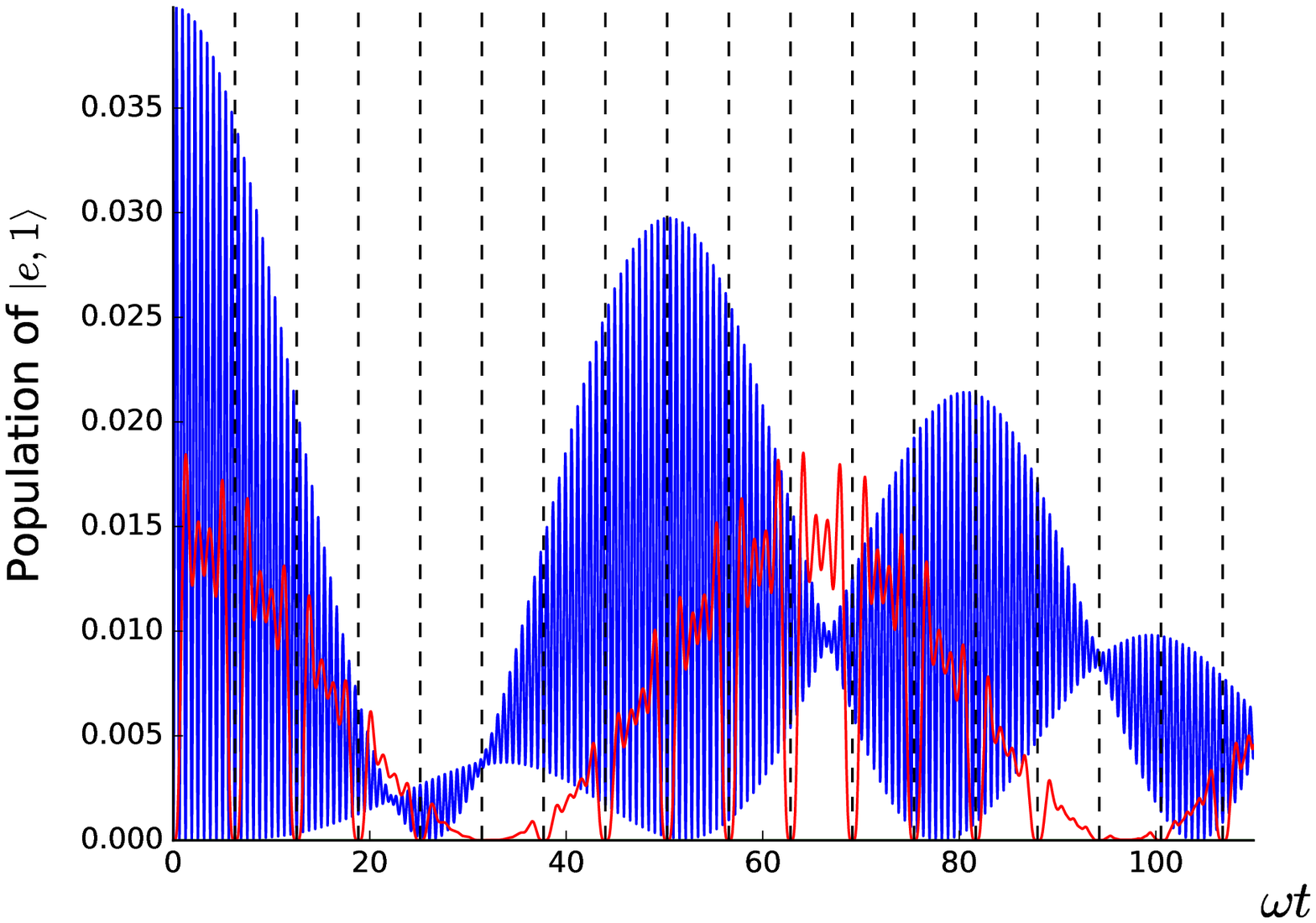}
\caption{\label{fig:e1_full_constraints}}
\end{subfigure}
\caption{\label{fig:full_constraints}Optimization under full effective Hamiltonian constraints. Populations (a) $P_{g1}(t)=\mbox{tr}(\ket{g,1}\bra{g,1}\rho(t))$ (b) $P_{e1}(t)=\mbox{tr}(\ket{e,1}\bra{e,1}\rho(t))$ for for the monochromatic dynamics (blue line), the target dynamics (green line) and the poly-chromatic dynamics (red line). The straight black dashed lines denote when $t$ is equal to the integral multiple of the period $T$. Parameters: $\text{f}_{tg}=0.1$, $m=10$, $\eta=0.05$ (Lamb-Dicke parameters are assumed to be the same for all frequencies) and $n=5$.}
\end{figure*}

As explained in subsection \ref{Numerical Optimization Of The Target Functional}, the minimization of the target functional is subject to the five constraints in Eq. \eqref{eq:constraints} instead of the possible choice of seven constraints. In Fig. \ref{fig:full_constraints}, the system is again initialised in an intial state $\rho_0=\ket{g,1}\bra{g,1}$, and the poly-chromatic pulse is generated using the full set of (seven) effective Hamiltonian constraints. Fig. \ref{fig:e1_full_constraints} shows that the carrier transition is again suppressed when comparing with monochromatic driving, however, the poly-chromatic pulse drives the system off resonance from the ideal red sideband process. We dicussed in subsection \ref{Periodic Control} that the poly-chromatically-driven dynamics should be extremely close to the target dynamics at $t=qT$ (where $q$ is an integer). The population of $\ket{e,1}$ under the new poly-chromatic pulse (Fig. \ref{fig:e1_full_constraints}) is still close to zero periodically. Meanwhile, the red sideband dynamics deviates from the target dynamics quite substantially even at $t=qT$. As a result, we choose to use Eq. \eqref{eq:constraints} to find the optimal pulse. Despite that the effective dynamics up to second order approximation gives accurate description to the actual dynamics as shown in Fig. \ref{fig:effective}, the use of the full set of (seven) constraints leaves less space for optimisation, which leads to the observed deterioration in performance of the optimal pulse.

\subsection{Stronger Driving and High-lying Motional States}\label{Stronger Driving and High-lying Motional States}

In previous subsections, we have demonstrated that the optimal poly-chromatic pulse performs better than the conventional monochromatic pulse. However, we have only tested the theory under weaking driving. In this subsection, we discuss how the optimal pulse performs beyond the weak driving limit.

In the weak driving limit,  we know from the Hamiltonian (Eq. \eqref{eq:simplified_hamiltonian}) that the red sideband transition scales with $\text{f}_{tg}$, the carrier transition scales with $\text{f}_{tg}/\eta m$ and the blue sideband transition scales with $\text{f}_{tg}/2m$. If we increase the target driving amplitude (\textit{i.e.} using the same parameters as in Fig. \ref{fig:k=1_weak} except that $\text{f}_{tg}$ is doubled in magnitude), both monochromatically- and poly-chromatically-driven dynmaics are then off-resonance from the target dynamics, since the carrier transition is no longer negligible under stronger driving. However, as the transition amplitude of the carrier transition scales with $\text{f}_{tg}/\eta m$, we can damp down the excitation by increasing the Lamb-Dicke parameter $\eta$ correspondingly, and this allows the possibility to drive the gate into faster dynamics provided that the value of $\eta$ is still within the Lamb-Dicke limit. 



The ideal red sideband dynamics oscillates between state $\ket{g,k}$ and state $\ket{e,k-1}$ with a Rabi-frequency of $\text{f}_{tg}\sqrt{k}/2$, and when comparing with low-lying motional states, high-lying motional states oscillate much quicker as if they are driven under strong laser fields. This implies that if the system begins with a high-lying motional state (say for example $\rho_0=\ket{g,15}\bra{g,15}$), then the poly-chromatic pulse would fail to keep driving the system on resonance in spite of the fact that the driving amplitude of the pulse is weak. With an intial state $\rho_0=\ket{g,15}\bra{g,15}$, the Rabi frequency is about four times larger than the one with an initial state $\rho_0=\ket{g,1}\bra{g,1}$. Under such circumstance, the system can no longer be considered as under weak driving, and the optimal pulse constructed under our current approximation cannot perform well since high order terms begin to have siginificant effects on the system. 


\subsection{State Infidelity vs. Gate Infidelity}\label{State Infidelity vs. Gate Infidelity}
\begin{figure}
\centering
\includegraphics[width=8.6cm]{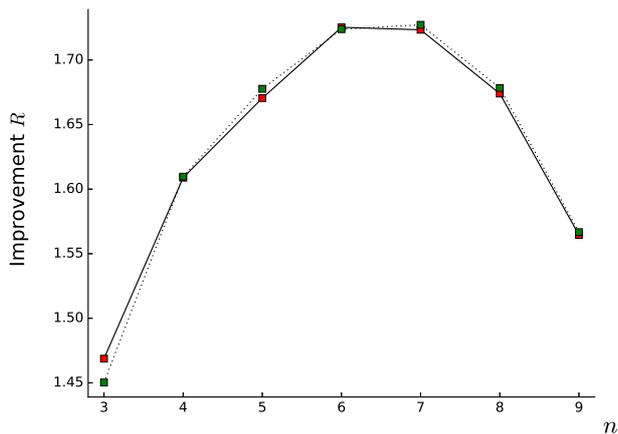}
\caption{\label{fig:state_vs_truncated_gate} State Infidelity vs. Truncated Gate Infidelity plotted as functions of $n$. The solid line (with red square) represents the performance of the optimal pulse found by minimizing the state infidelity (Eq. \eqref{eq:state_infidelity_simplified}), while the dotted line (with green square) represents the performance of the optimal pulse found by minimizing the truncated gate infidelity (Eq. \eqref{eq:truncated_gate_infidelity}).  Parameters: $\text{f}_{tg}=0.1$, $m=10$, $\eta=0.05$ (Lamb-Dicke parameters are assumed to be the same for all frequencies) and $d=2$.}
\end{figure}

We have derived the state infidelity (Eq. \eqref{eq:state_infidelity_simplified}) and the asymptotic gate infidelity (Eq. \eqref{eq:gate_infidelity_asymptotic}) respectively, and in previous subsections, we have used the state infidelity to determine the optimal driving pulse. In this subsection, we discuss how the optimal pulse found by miniziming the gate infidelity differs from the pulse found by minimizing the state infidelity.

The constructed poly-chromatic pulse cannot maintain good control of the sideband transition if the system is initialized in a high-lying motional state. Meanwhile, the asymptotic gate infidelity (Eq. \eqref{eq:gate_infidelity_asymptotic}) ignores low-lying motional states, the optimal pulse found by miniziming Eq. \eqref{eq:gate_infidelity_asymptotic} is thus not expected to simulate the target dynamics as well as the optimal pulse found by miniziming Eq. \eqref{eq:state_infidelity_simplified}. Since the gate infidelity (Eq. \eqref{eq:gate_infidelity_simplified}) does not return finite values for an infinite-dimensional system, we instead truncate the system to a finite-dimensional subspace spanned by $\{\ket{g,0},\ket{e,0},\ldots,\ket{g,d-1},\ket{e,d-1}\}$ and replace the upper limit of the summation in Eq. \eqref{eq:gate_infidelity_simplified} by $d-1$, where $d-1$ denotes the highest motional number in the subspace. As a result, the truncated gate infidelity reads
\begin{equation}\label{eq:truncated_gate_infidelity}
\mathcal{I}_{gate,tr}=\frac{1}{T}\int^{T}_0 dt \, \Big(2d|g_1|^2+d^2|g_2|^2+d^2|g_3|^2\Big)
\end{equation}
The way to choose the motional number $d$ seems arbitrary at first, but based on the findings of previous subsections, $d$ should be small such that the subspace covers only the relevant low-lying motional states.

As we discussed in subsection \ref{Comparison With Monochromatic Driving}, the actual performance during (at least) one gate cycle cannot be reflected by Eq. \eqref{eq:state_infidelity_simplified}, and instead we defined Eq. \eqref{eq:infidelity_cycle} to measure how well the pulse can drive the system during one cycle of gate operation. In this subsection, we continue to use the same definitions of both $\mathcal{I}_{cycle}$ and improvement $R$ to compare optimal pulses found by the state infidelity and the truncated gate infidelity respectively. In Fig. \ref{fig:state_vs_truncated_gate}, the system is initialised in an initial state $\rho_0=\ket{g,1}\bra{g,1}$. The solid line in Fig. \ref{fig:state_vs_truncated_gate} shows the performance of the optimal pulse found by the state infidelity under one cycle (same line shown in Fig. \ref{fig:improvement}), while the dotted line in Fig. \ref{fig:state_vs_truncated_gate} shows the performance of the optimal pulse found by the truncated gate infidelity under one cycle (where $d$ is arbitrarily chosen to be equal to 2). The pulse determined by Eq. \eqref{eq:truncated_gate_infidelity} in Fig. \ref{fig:state_vs_truncated_gate} optimizes the dynamics of the subspace spanned by $\{\ket{g,0},\ket{e,0},\ket{g,1},\ket{e,1}\}$ (which are the most relevant states at low temperature), and furthermore, Eq. \eqref{eq:truncated_gate_infidelity} shares a similar mathematical structure to Eq. \eqref{eq:state_infidelity_simplified}. As a result,  both pulses determined by Eq. \eqref{eq:truncated_gate_infidelity} and Eq. \eqref{eq:state_infidelity_simplified} perform equally well (see Fig. \ref{fig:state_vs_truncated_gate}).

\begin{figure}
\centering
\includegraphics[width=8.6cm]{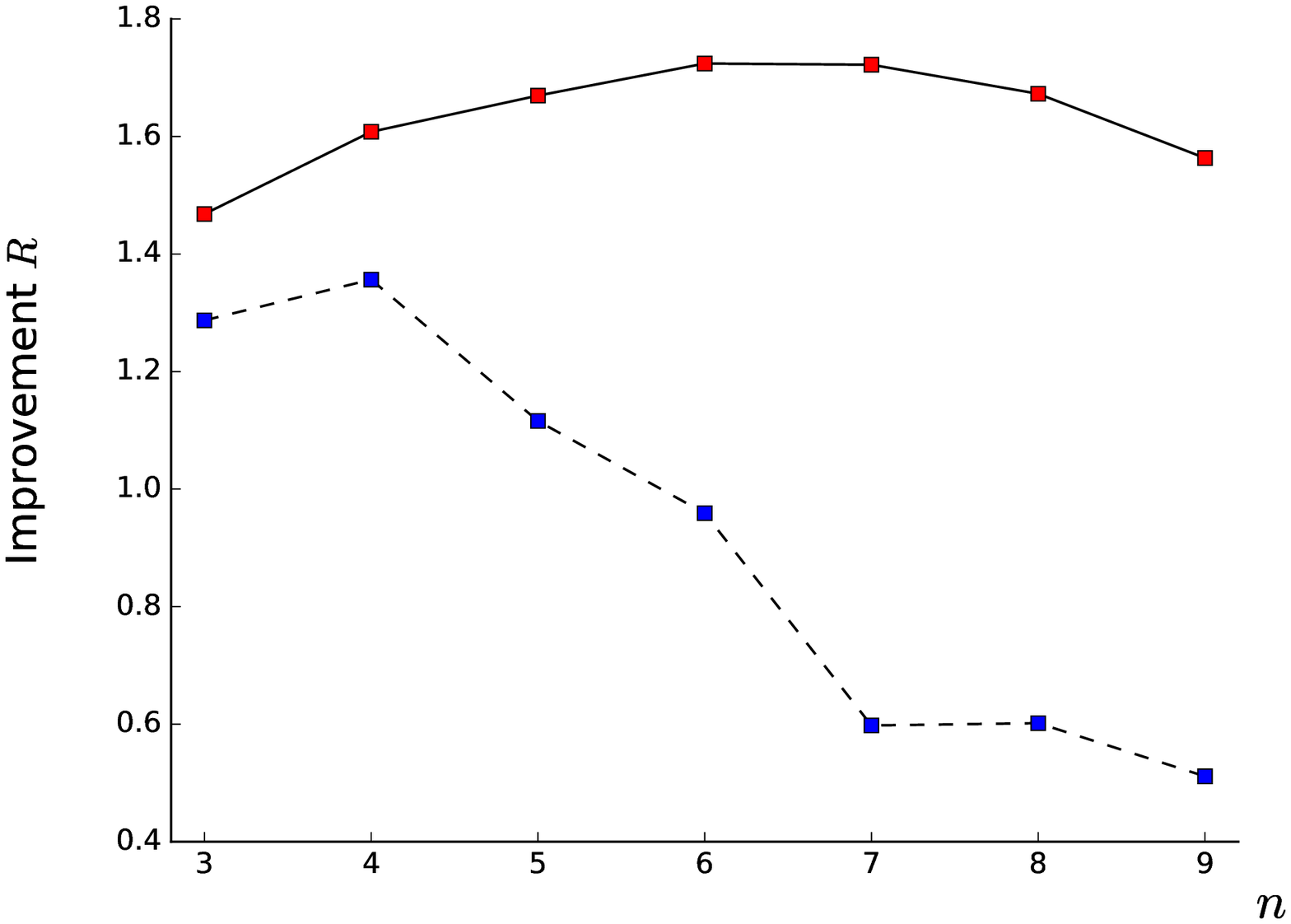}
\caption{\label{fig:state_vs_asymptotic_gate} State Infidelity vs. Asymptotic Gate Infidelity plotted as functions of $n$. The solid line (with red square) represents the performance of the optimal pulse found by minimizing the state infidelity (Eq. \eqref{eq:state_infidelity_simplified}), while the dahed line (with blue square) represents the performance of the optimal pulse found by minimizing the asymptotic gate infidelity (Eq. \eqref{eq:gate_infidelity_asymptotic}).  Parameters: $\text{f}_{tg}=0.1$, $m=10$ and $\eta=0.05$ (Lamb-Dicke parameters are assumed to be the same for all frequencies).}
\end{figure}

Earlier, we argue that the optimal pulse found by miniziming the asymptotic gate infidelity is not expected to perform well, and we provide here Fig. \ref{fig:state_vs_asymptotic_gate} to support the argument. From the plot, the pulse determined by the state infidelity clearly performs better. The asymptotic gate infidelity (Eq. \eqref{eq:state_infidelity_simplified}) is independent of the fluctuations in the carrier transition, which implies that the pulse determined by the gate infidelity does not minimize the fluctuations in one of the undesired transitions. The fluctuations in the carrier transition within each period are ignored in the asymptotic expression, and as a result, the optimal pulse determined by Eq. \eqref{eq:state_infidelity_simplified} performs poorly. Furthermore, the poly-chromatic pulse contains slow frequency terms which are not present in the monochromatic pulse, and along with the fact that the fluctuations in the carrier transition are absent in the asymptotic expression (Eq. \eqref{eq:state_infidelity_simplified}), the pulse determined by Eq. \eqref{eq:state_infidelity_simplified} performs even worse than the monochromatic pulse for $n=6, 7, 8, 9$ (\textit{i.e.} $R$ is below unity).

By comparing the three different definitions of a target functional (Eq. \eqref{eq:gate_infidelity_asymptotic}, Eq. \eqref{eq:state_infidelity_simplified} and Eq. \eqref{eq:truncated_gate_infidelity} respectively), we identify that the asymptotic expression (Eq. \eqref{eq:gate_infidelity_asymptotic}) is not a good choice since the fluctuations in the carrier transition are completely ignored. In Fig. \ref{fig:state_vs_truncated_gate}, we show that optimal pulse determined by either Eq. \eqref{eq:state_infidelity_simplified} or Eq. \eqref{eq:truncated_gate_infidelity} can improve the simulation of the target dynamics. Overall, the choice of using either Eq. \eqref{eq:state_infidelity_simplified} or Eq. \eqref{eq:truncated_gate_infidelity} as the definition of a target functional is equally good and depends on the reader's preference.

\section{Summary and Outlook}\label{Summary and Outlook}

As hardware for quantum information processing is being scaled up, the number of available transitions that couple to a driving field grows.
Together with the need of realising high-fidelity operations on time-scales that are short as compared to the coherence time, the selection of resonances as sole basis of control poses severe limitations.
Suitably shaped polychromatic driving offers many solutions to improve gate fidelities and to avoid the impact of undesired processes \cite{haddadfarshi2016high}.
The case considered here with a transition that couples more weakly to the driving field than undesired transitions allows us to explore the limitations of practical control schemes.
As we have seen, there is substantial room for improvement in the regime of low temperatures, but as the range of potential initial states and the driving strength grows, the added value of the present control techniques becomes smaller.
This is likely a result of the approximations that the largely analytic framework relies on.
Such limitations might be overcome with numerical techniques \cite{khaneja2005optimal,PhysRevLett.115.190801,PhysRevLett.106.190501} that are better suited for strong driving.
In contrast to the present analytic techniques, numerical approaches necessarily suffer from the unfavourable scaling of composite quantum systems,
and if system size and required driving strength limit the usefulness of either approach,
recently developed strategies based on experimental design of control \cite{2017arXiv170500565B,2017arXiv170101198L,2017arXiv170101723D} offer a solution.

\onecolumngrid
\appendix
\section{Effective Hamiltonian and Constraints in the Minimization} \label{appendix:Effective Hamiltonian}

In the main text, we have made two assumptions about the system Hamiltonian (Eq.~\eqref{eq:simplified_hamiltonian} and Eq. \eqref{eq:time_dependent_terms}):  (1) The Hamiltonian is $T$-periodic  (2) The driving amplitude ($\text{f}_{tg}\omega$) of the pulse is sufficiently small compared to $\omega$. Under the assumptions of periodicity and weak driving, the dynamics of the system can be understood in terms of a time-independent effective Hamiltonian. In this section, we provide the first three terms of the effective Hamiltonian.

We begin here with the lowest order effective term. By applying the Magnus formula in Eq. \eqref{eq:magnus_terms}, the zeroth term of the effective Hamiltonian $H_{eff}^{(0)}$ reads,
\begin{equation}\label{eq:0th_eff}
\begin{split}
H_{eff}^{(0)}&=\frac{-\delta \omega}{4} \sigma_{z} +\frac{i\eta_{0}f_{0} \omega}{2}\sigma_+ a+h.c.\\
&=\alpha_{1}^{(0)}\sigma_z+\alpha_{2}^{(0)}\sigma_+a+h.c.
\end{split}
\end{equation}
The above $H_{eff}^{(0)}$ has two terms, the first term ($\sigma_z$) is a Lamb-shift term in this interaction frame and the second term ($\sigma_+a$) is responsible for the red sideband transition at this order of approximation. Under sufficiently weak driving, the zeroth order effective dynamics is in principle the same as the driven dynamics for a sufficiently short period of time. However, higher order terms begin to affect the dynamics of the system significantly on a long time scale. Thus, the calculation must go beyond the zeroth order.

In order to calculate the first order term of the effective Hamiltonian $H_{eff}^{(1)}$, the commutator of the Hamiltonian (Eq. \eqref{eq:simplified_hamiltonian}) at time $t_1$ and time $t_2$ needs to be evaluated.
\begin{equation}\label{eq:commutator}
\begin{split}
&[\bar{H}(t_1),\bar{H}(t_2)]\\
=&-\delta \omega\big((h_1(t_2)-h_1(t_1))\sigma_{+}+(h_2(t_2)-h_2(t_1))\sigma_{+}a+(h_3(t_2)-h_3(t_1))\sigma_{+}a^{\dagger}\big)+h_1(t_1)h_1^*(t_2)\sigma_z\\
&+h_1(t_1)h_2^*(t_2)\sigma_za^{\dagger}+h_1(t_1)h_3^*(t_2)\sigma_za+h_2(t_1)h_1^*(t_2)\sigma_za+h_2(t_1)h_2^*(t_2)(\sigma_zaa^{\dagger}+\sigma_-\sigma_+)\\
&+h_2(t_1)h_3^*(t_2)\sigma_za^2+h_3(t_1)h_1^*(t_2)\sigma_za^{\dagger}+h_3(t_1)h_2^*(t_2)\sigma_z(a^{\dagger})^2+h_3(t_1)h_3^*(t_2)(\sigma_z a^{\dagger}a-\sigma_- \sigma_+)-h.c.\\
\end{split}
\end{equation}
In the evaluation of $H_{eff}^{(1)}$, one needs to integrate exponential terms such as $e^{i(q-j-m)\omega t}$. For the case ($m>2n$), the integral $\int_0^T dt\,e^{i(q-j-m)\omega t}=0$; while for the case ($2n\geq m>n$), the integral $\int_0^T dt\,e^{i(q-j-m)\omega t}$ is not equal to zero if $q-j-m=0$. Furthermore, two more cases are possible in the evaluation of  $H_{eff}^{(1)}$:  (1) 2m>$n\geq m$  (2) $n\geq 2m$. There are two reasons why these two cases are not investigated in this paper:  (1) $n$ is chosen not to be too large to avoid unrealistic experimental implementation  (2) Similar approaches to improve quantum simulations \cite{verdeny2014optimal,haddadfarshi2016high} show that the increase in number of frequencies only improves the gate performance asymptotically. Overall, $H_{eff}^{(1)}$ can be expressed in shorthand notations,
\begin{equation}\label{eq:shorthand_eff_1}
\begin{split}
H_{eff}^{(1)}=&\alpha_{1}^{(1)}\sigma_++\alpha_{2}^{(1)}\sigma_+a+\alpha_{3}^{(1)}\sigma_+a^{\dagger}+\alpha_{4}^{(1)}\sigma_z+\alpha_{5}^{(1)}\sigma_za+\alpha_{6}^{(1)}\sigma_za^{2}\\
&+\alpha_{7}^{(1)}(\sigma_zaa^{\dagger}+\sigma_-\sigma_+)+\alpha_{8}^{(1)}(\sigma_za^{\dagger}a-\sigma_-\sigma_+)+h.c.
\end{split}
\end{equation}
where the shorthand notations $\alpha_{j}^{(1)}$ are introduced to simplify the above Eq. \eqref{eq:shorthand_eff_1} and their analytical expressions are provided in Appendix \ref{appendix:First Order Effective Hamiltonian}. The first three terms in $H_{eff}^{(1)}$ induce the carrier, the redsidband and the bluesideband transitions respectively. The fourth ($\sigma_z$), seventh ($\sigma_zaa^{\dagger}+\sigma_-\sigma_+$) and eighth ($\sigma_za^{\dagger}a-\sigma_-\sigma_+$) terms do not cause any quantum transitions, but these three effective terms cause shifts to the energy levels of the unperturbed system. The fifth term ($\sigma_za^{\dagger}$) and its conjugate induce transitions between motional states $\ket{k-1}$ and $\ket{k}$, while the sixth term ($\sigma_za^2$) and its conjugate induce transitions between motional states $\ket{k-2}$ and $\ket{k}$.

\begin{figure*}
\begin{subfigure}{0.5\textwidth}
\centering
\includegraphics[width=8.6cm]{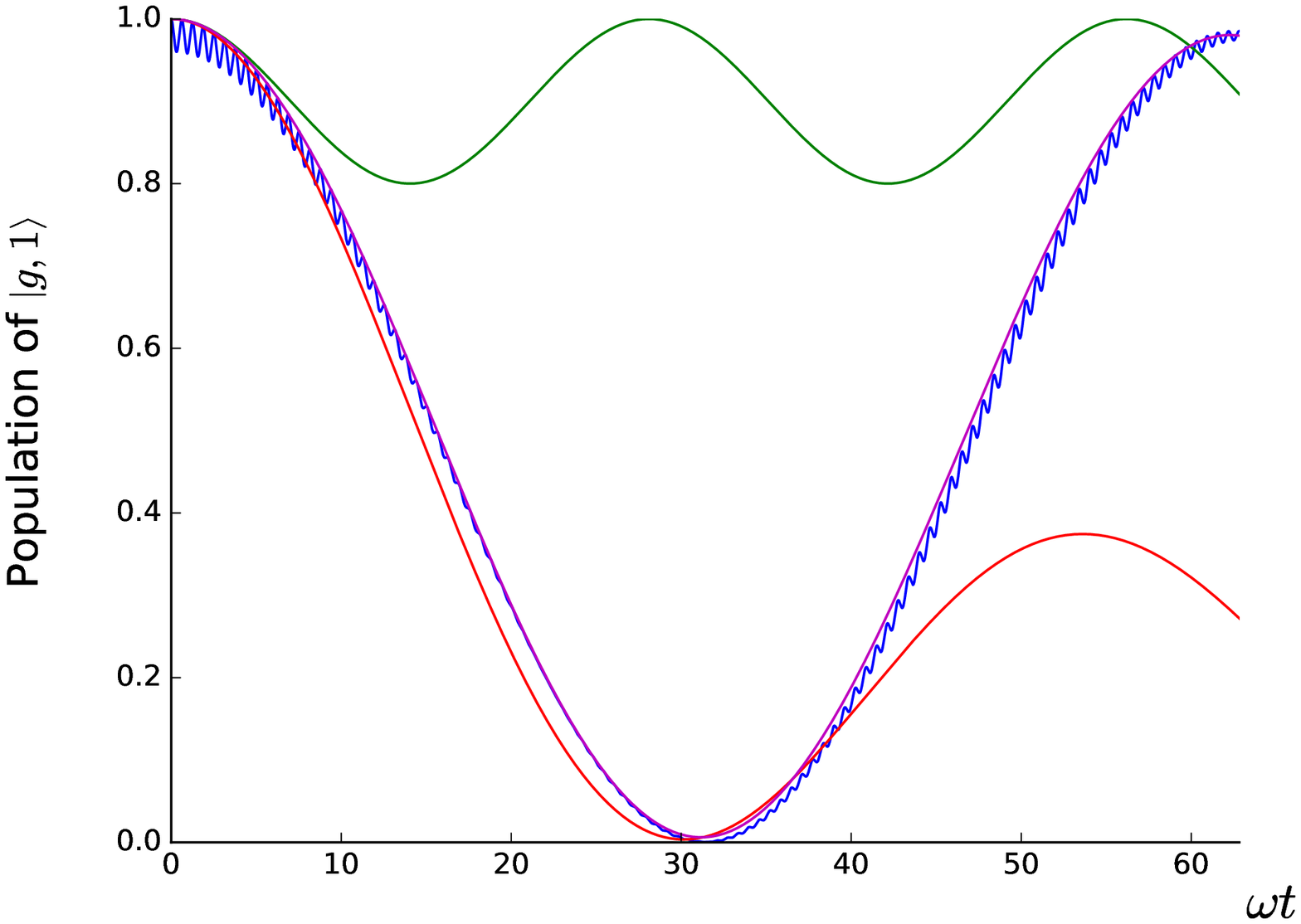} 
\caption{\label{fig:effective_g1}}
\end{subfigure}\hfill
\begin{subfigure}{0.5\textwidth}
\centering
\includegraphics[width=8.6cm]{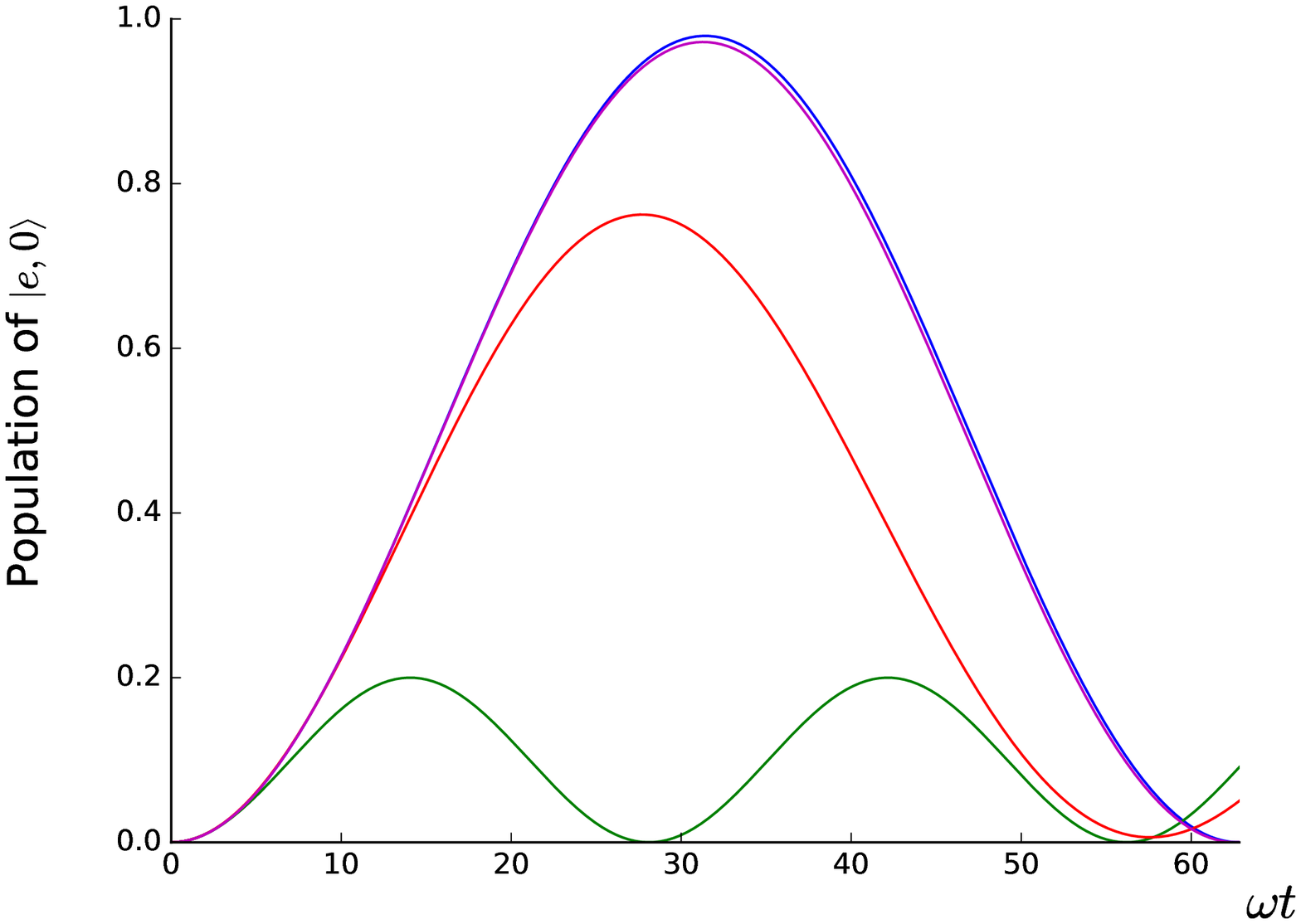}
\caption{\label{fig:effective_e0}}
\end{subfigure}
\caption{\label{fig:effective} Effective dynamics vs. Actual driven dynamics under monochromatic driving ($n=0$). Populations (a) $P_{g1}(t)=\mbox{tr}(\ket{g,1}\bra{g,1}\rho(t))$ (b) $P_{e0}(t)=\mbox{tr}(\ket{e,0}\bra{e,0}\rho(t))$ induced by the zeroth order effective Hamiltonian (green line), the first order effective Hamiltonian (red line), the second order effective Hamiltonian (magenta line) and the actual Hamitlonian (blue line) respectively. Parameters: $\text{f}_{tg}=0.1$, $m=10$, $\eta=0.05$, $\delta=0.2$ and $\text{f}_0=2$.}
\end{figure*}

To further improve the approximation of the effective dynamics, the second order term of the effective Hamiltonian also needs to be evaluated. In the evaluation of $H_{eff}^{(2)}$, the nested commutator $[\bar{H}(t_{i_1}),[\bar{H}(t_{i_2}),\bar{H}(t_{i_3})]]$ needs to be determined, where the details of the derivation are shown in Appendix \ref{appendix:Second Order Effective Hamiltonian}. Using short hand notations, we express $H_{eff}^{(2)}$ in the following way,
\begin{equation}\label{eq:shorthand_eff_2}
\begin{split}
H_{eff}^{(2)}=&\big(\alpha_{1}^{(2)}+\alpha_{9}^{(2)}\big)\sigma_++\big(\alpha_{2}^{(2)}+\alpha_{10}^{(2)}\big)\sigma_+a+\big(\alpha_{3}^{(2)}+\alpha_{11}^{(2)}\big)\sigma_+a^{\dagger}+\alpha_{4}^{(2)}\sigma_z+\alpha_{5}^{(2)}\sigma_za\\
&+\alpha_{6}^{(2)}\sigma_za^{2}+\alpha_{7}^{(2)}(\sigma_zaa^{\dagger}+\sigma_-\sigma_+)+\alpha_{8}^{(2)}(\sigma_za^{\dagger}a-\sigma_-\sigma_+)+h.c.+O(\eta^2,\eta^3)
\end{split}
\end{equation}
where the lengthy closed-form expressions of terms such as $\alpha_{1}^{(2)}$ are provided in Appendix \ref{appendix:List of Integrals}. As explained in subsection \ref{Poly-Chromatic Driving Scheme}, we have assumed that the system is cooled to the Lamb-Dicke regime (informally $\eta\ll1$). To avoid further complication, we choose to neglect all the terms assocaited with $\eta^2$ or $\eta^3$ in the third order Magnus calculation (see Appendix \ref{appendix:Second Order Effective Hamiltonian}). Both Eq. \eqref{eq:shorthand_eff_1} and Eq. \eqref{eq:shorthand_eff_2} share the same mathematical structure, which implies that the second order term $H_{eff}^{(2)}$  induces the same type of transitions as the first order term $H_{eff}^{(1)}$ does. Overall, the effective Hamiltonian takes the following form,
\begin{equation}\label{eq:explicit_eff_hamiltonian}
\begin{split}
H_{eff}\approx&H_{eff}^{(0)}+H_{eff}^{(1)}+H_{eff}^{(2)}\\
\approx& \big(\alpha_{1}^{(1)}+\alpha_{1}^{(2)}+\alpha_{9}^{(2)}\big)\sigma_++\big(\alpha_{2}^{(0)}+\alpha_{2}^{(1)}+\alpha_{2}^{(2)}+\alpha_{10}^{(2)}\big)\sigma_+a+\big(\alpha_{3}^{(1)}+\alpha_{3}^{(2)}+\alpha_{11}^{(2)}\big)\sigma_+a^{\dagger}\\
&+\big(\alpha_{1}^{(0)}+\alpha_{4}^{(1)}+\alpha_{4}^{(2)}\big)\sigma_z+\big(\alpha_{5}^{(1)}+\alpha_{5}^{(2)})\sigma_za+(\alpha_{6}^{(1)}+\alpha_{6}^{(2)}\big)\sigma_za^{2}+\big(\alpha_{7}^{(1)}+\alpha_{7}^{(2)}\big)(\sigma_zaa^{\dagger}+\sigma_-\sigma_+)\\
&+\big(\alpha_{8}^{(1)}+\alpha_{8}^{(2)}\big)(\sigma_za^{\dagger}a-\sigma_-\sigma_+)+h.c.+O(\eta^2,\eta^3) 
\end{split}
\end{equation}
and making use of the two identities, $aa^{\dagger}=a^{\dagger}a+\mathds{1}$ and $\sigma_-\sigma_+=(\mathds{1}-\sigma_z)/2$, we can then write the effective Hamiltonian into a sum of independent operators,
\begin{equation}\label{eq:explicit_eff_hamiltonian_full}
\begin{split}
H_{eff}=&\frac{1}{2}\big(\alpha_{7}^{(1)}+\alpha_{7}^{(2)}-\alpha_{8}^{(1)}-\alpha_{8}^{(2)}\big)\mathds{1}+\big(\alpha_{1}^{(0)}+\alpha_{4}^{(1)}+\alpha_{4}^{(2)}+\frac{1}{2}(\alpha_{7}^{(1)}+\alpha_{7}^{(2)}+\alpha_{8}^{(1)}+\alpha_{8}^{(2)})\big)\sigma_z\\
&+\big(\alpha_{5}^{(1)}+\alpha_{5}^{(2)}\big)\sigma_za+\big(\alpha_{6}^{(1)}+\alpha_{6}^{(2)}\big)\sigma_za^{2}+\big(\alpha_{7}^{(1)}+\alpha_{7}^{(2)}+\alpha_{8}^{(1)}+\alpha_{8}^{(2)}\big)\sigma_za^{\dagger}a\\
&+ \big(\alpha_{1}^{(1)}+\alpha_{1}^{(2)}+\alpha_{9}^{(2)}\big)\sigma_++\big(\alpha_{2}^{(0)}+\alpha_{2}^{(1)}+\alpha_{2}^{(2)}+\alpha_{10}^{(2)}\big)\sigma_+a+\big(\alpha_{3}^{(1)}+\alpha_{3}^{(2)}+\alpha_{11}^{(2)}\big)\sigma_+a^{\dagger} \hspace{1.49cm}\\
&+h.c.+O(\eta^2,\eta^3) 
\end{split}
\end{equation}
where the above Eq. \eqref{eq:explicit_eff_hamiltonian_full} shows how to evaluate the coefficients ($c_i$) in Eq. \eqref{eq: effective_ham} in the main text. 

In Fig. \ref{fig:effective}, the system is initialised in an intial state $\rho_0=\ket{g,1}\bra{g,1}$. Fig. \ref{fig:effective_g1} and Fig. \ref{fig:effective_e0} depict the population of $\ket{g,1}$ and $\ket{e,0}$ respectively. The evolution of the system shown in Fig. \ref{fig:effective} is driven by four different Hamiltonians: (1) Zeroth order effective Hamiltonian  (2) First order effective Hamiltonian  (3) Second order effective Hamiltonian  (4) Actual Hamiltonian under monochromatic laser field. As can be seen from Fig. \ref{fig:effective}, the second order effective Hamiltonian gives the best approximation to the dynamics driven by the actual Hamiltonian (Eq. \ref{eq:simplified_hamiltonian}).

In section \ref{Numerical Results}, we have discussed that the poly-chromatic pulse performs worse than the monochromatic pulse if we impose all the seven constraints from the effective Hamiltonian. Instead, we have found that the optimal combination of constraints should be
\begin{equation}\label{eq:constraints}
\begin{split}
\alpha_{5}^{(1)}+\alpha_{5}^{(2)}&=0\\
\alpha_{6}^{(1)}+\alpha_{6}^{(2)}&=0\\
\alpha_{1}^{(0)}+\alpha_{4}^{(1)}+\alpha_{4}^{(2)}&=0\\
\alpha_{7}^{(1)}+\alpha_{7}^{(2)}+\alpha_{8}^{(1)}+\alpha_{8}^{(2)}&=0\\
\alpha_{2}^{(0)}+\alpha_{2}^{(1)}+\alpha_{10}^{(2)}&=\frac{i\text{f}_{tg}\omega}{2}\\
\end{split}
\end{equation}
where readers might notice that $\alpha_{2}^{(2)}$ is not included in the last constraint of Eq. \eqref{eq:constraints}. As we have found that if $\alpha_{2}^{(2)}$ is included, the minimization result again leads to a smaller effective Rabi-frequency when compared to the Rabi-frequency of the target dynamics. As a consequence, $\alpha_{2}^{(2)}$ is not included in the minimization of the target functional.

\section{First Order Term of the Effective Hamiltonian} \label{appendix:First Order Effective Hamiltonian}
In Appendix \ref{appendix:Effective Hamiltonian}, we discussed the general aspect of the theory of effective Hamiltonian, and in this part of the paper, we provide the analytical expression of the first order term of the effective Hamiltonian. The commutator of the Hamiltonian (Eq. \eqref{eq:simplified_hamiltonian}) at time $t_1$ and time $t_2$ are provided in Eq. \eqref{eq:commutator}, and by integrating the coefficients ($\alpha_{j}^{(1)}$) in Eq. \eqref{eq:commutator}, the first order term of the effective Hamiltonian can then be determined. In the following, we provide the analytical expressions of $\alpha_{j}^{(1)}$ where $m>n$.
\\
\begin{equation}
\begin{split}
&\alpha_{1}^{(1)}=\frac{-i}{2T} \int_{0}^{T}dt_{1}\int_{0}^{t_1}dt_2 \, \Big(\delta \omega (h_1(t_1)-h_1(t_2)) \Big)\hspace{9.25cm}\\
&\text{For}\,\, m>n:\\
&\hspace{0.605cm}=\sum_{j=-n}^{n} \frac{\text{f}_j \delta \omega}{2(m-j)}\\
\end{split}
\end{equation}
\begin{equation}
\begin{split}
&\alpha_{2}^{(1)}=\frac{-i}{2T} \int_{0}^{T}dt_{1}\int_{0}^{t_1}dt_2 \, \Big(\delta \omega (h_2(t_1)-h_2(t_2)) \Big)\hspace{9.25cm}\\
&\text{For}\,\, m>n:\\
&\hspace{0.605cm}=\sum_{\substack{j=-n\\(j \neq 0)}}^{n} \frac{i \eta_j\text{f}_j \delta \omega}{2j}\\
\end{split}
\end{equation}
\begin{equation}
\begin{split}
&\alpha_{3}^{(1)}=\frac{-i}{2T} \int_{0}^{T}dt_{1}\int_{0}^{t_1}dt_2 \, \Big(\delta \omega (h_3(t_1)-h_3(t_2)) \Big)\hspace{9.25cm}\\
&\text{For}\,\, m>n:\\
&\hspace{0.605cm}=-\sum_{j=-n}^{n} \frac{i\eta_j \text{f}_j \delta \omega}{2(2m-j)} \\
\end{split}
\end{equation}
\begin{equation}
\begin{split}
&\alpha_{4}^{(1)}=\frac{-i}{4T} \int_{0}^{T}dt_{1}\int_{0}^{t_1}dt_2 \,\Bigg(h_1(t_1)h_1^*(t_2)-c.c.\Bigg) \hspace{9.28cm}\\
&\text{For}\,\, m>n:\\
&\hspace{0.605cm}=\sum_{j=-n}^{n} \frac{(\text{f}_j)^2 \omega}{8(m-j)} \\
\end{split}
\end{equation}
\begin{equation}
\begin{split}
&\alpha_{5}^{(1)}=\frac{-i}{2T} \int_{0}^{T}dt_{1}\int_{0}^{t_1}dt_2 \, \Big(-h_1^*(t_1)h_2(t_2)+h_1(t_1)h_3^*(t_2)+h_2(t_1)h_1^*(t_2)-h_3^*(t_1)h_1(t_2)\Big) \hspace{3.075cm}\\
&\text{For}\,\, m>2n:\\
&\hspace{0.605cm}=-\sum_{j=-n}^{n} \frac{i\eta_0 \text{f}_0 \text{f}_j \omega}{4(m-j)} \\
&\text{For}\,\, 2n\geq m>n:\\
&\hspace{0.605cm}=-\Bigg(\sum_{j=-n}^{n} \frac{i\eta_0 \text{f}_0 \text{f}_j \omega}{4(m-j)}-\sum_{\substack{j,q=-n\\(m+j-q= 0)}}^{n} \frac{i\eta_j \text{f}_j \text{f}_q \omega}{4(m-q)}+\sum_{\substack{j,q=-n\\(m+q-j= 0)}}^{n} \frac{i\eta_j \text{f}_j \text{f}_q \omega}{4(2m-j)}\Bigg) 
\end{split}
\end{equation}
\begin{equation}
\begin{split}
&\alpha_{6}^{(1)}=\frac{-i}{2T} \int_{0}^{T}dt_{1}\int_{0}^{t_1}dt_2 \, \Big(h_2(t_1)h_3^*(t_2)-h_3^*(t_1)h_2(t_2)\Big) \hspace{8.025cm}\\
&\text{For}\,\, m>n:\\
&\hspace{0.605cm}=-\sum_{j=-n}^{n} \frac{\eta_0 \text{f}_0 \eta_j \text{f}_j \omega}{4(2m-j)}\\
\end{split}
\end{equation}
\begin{equation}
\begin{split}
&\alpha_{7}^{(1)}=\frac{-i}{4T} \int_{0}^{T}dt_{1}\int_{0}^{t_1}dt_2 \,\Bigg(h_2(t_1)h_2^*(t_2)-c.c.\Bigg) \hspace{9.28cm}\\
&\text{For}\,\, m>n:\\
&\hspace{0.605cm}=\Bigg(\sum_{\substack{j=-n\\(j \neq 0)}}^{n} \frac{\eta_0\text{f}_0 \eta_j\text{f}_j \omega}{4j}-\sum_{\substack{j=-n\\(j \neq 0)}}^{n} \frac{(\eta_j\text{f}_j)^2 \omega}{8j}\Bigg)\\
\end{split}
\end{equation}
\begin{equation}
\begin{split}
&\alpha_{8}^{(1)}=\frac{-i}{4T} \int_{0}^{T}dt_{1}\int_{0}^{t_1}dt_2 \,\Bigg(h_3(t_1)h_3^*(t_2)-c.c.\Bigg) \hspace{9.28cm}\\
&\text{For}\,\, m>n:\\
&\hspace{0.605cm}=\sum_{j=-n}^{n} \frac{(\eta_j \text{f}_j)^2 \omega}{8(2m-j)}\\
\end{split}
\end{equation}
As explained in subsection \ref{appendix:Effective Hamiltonian}, we are only interested in the above two cases ($m>2n$ \& $2n\geq m>n$); while the two other possible cases ($2m>n\geq m$ \& $n\geq 2m$) are not investigated in this paper, but could easily be evaluated by following the same procedure if readers are interested.

\newpage
\section{Second Order Term of the Effective Hamiltonian}\label{appendix:Second Order Effective Hamiltonian}
We provide here the details of the evaluation of the second order term of the effective Hamiltonian. The second order term of the effective Hamiltonian is determined by the third order Magnus term, and as a consequence, the nested commutator $[\bar{H}(t_{1}),[\bar{H}(t_{2}),\bar{H}(t_{3})]]+[\bar{H}(t_{3}),[\bar{H}(t_{2}),\bar{H}(t_{1})]]$ needs to be determined:
\begin{equation}\label{eq:nested commutator}
\begin{split}
&[\bar{H}(t_1),[\bar{H}(t_2),\bar{H}(t_3)]]+[\bar{H}(t_3),[\bar{H}(t_2),\bar{H}(t_1)]]\\
=&+\Big((\delta \omega)^2 h_1(t_3)+(\delta \omega)^2 h_1(t_1)-2(\delta \omega)^2 h_1(t_2)\Big)\sigma_+\\
&+\Big((\delta \omega)^2 h_2(t_3)+(\delta \omega)^2 h_2(t_1) -2(\delta \omega)^2 h_2(t_2)\Big)\sigma_+a\\
&+\Big((\delta \omega)^2 h_3(t_3)+(\delta \omega)^2 h_3(t_1)-2(\delta \omega)^2 h_3(t_2)\Big) \sigma_+a^{\dagger}\\
&+\Big(\delta \omega h_1^*(t_1)h_1(t_3)+\delta \omega h_1(t_1) h_1^*(t_3)-\delta \omega h_1^*(t_1)h_1(t_2)-\delta \omega h_1(t_2)h_1^*(t_3)\Big)\sigma_z\\
&+\Big(\delta \omega h_1^*(t_1)h_2(t_3)+\delta \omega h_2(t_1) h_1^*(t_3)-\delta \omega h_1^*(t_1)h_2(t_2)-\delta \omega h_2(t_2)h_1^*(t_3)\\
&+\delta \omega h_1(t_1)h_3^*(t_3)+\delta \omega h_3^*(t_1)h_1(t_3)-\delta \omega h_1(t_1)h_3^*(t_2)-\delta \omega h_3^*(t_2)h_1(t_3)\\
&+\delta \omega h_2(t_1)h_1^*(t_3)+\delta \omega h_1^*(t_1)h_2(t_3)-\delta \omega h_2(t_1)h_1^*(t_2)-\delta \omega h_1^*(t_2)h_2(t_3)\\
&+\delta \omega h_3^*(t_1)h_1(t_3)+\delta \omega h_1(t_1)h_3^*(t_3)-\delta \omega h_3^*(t_1)h_1(t_2)-\delta \omega h_1(t_2)h_3^*(t_3)\Big)\sigma_za\\
&+\Big(\delta \omega h_2(t_1)h_3^*(t_3)+\delta \omega h_3^*(t_1)h_2(t_3)-\delta \omega h_2(t_1)h_3^*(t_2)-\delta \omega h_3^*(t_2)h_2(t_3)\\
&+\delta \omega h_3^*(t_1)h_2(t_3)+\delta \omega h_2(t_1)h_3^*(t_3)-\delta \omega h_3^*(t_1)h_2(t_2)-\delta \omega h_2(t_2)h_3^*(t_3)\Big)\sigma_za^2\\
&+\Big(\delta \omega h_2^*(t_1)h_2(t_3)+\delta \omega h_2(t_1)h_2^*(t_3)-\delta \omega h_2^*(t_1)h_2(t_2)-\delta \omega h_2(t_2)h_2^*(t_3)\Big)(\sigma_zaa^{\dagger}+\sigma_-\sigma_+)\\
&+\Big(\delta \omega h_3^*(t_1)h_3(t_3)+\delta \omega h_3(t_1) h_3^*(t_3)-\delta \omega h_3^*(t_1)h_3(t_2)-\delta \omega h_3(t_2)h_3^*(t_3)\Big)(\sigma_za^{\dagger}a-\sigma_-\sigma_+)\\
&+\Big(-2h_1^*(t_1)h_1(t_2)h_1(t_3)+4h_1(t_1)h_1^*(t_2)h_1(t_3)-2h_1(t_1)h_1(t_2)h_1^*(t_3)\Big)\sigma_+\\
&+\Big(-2h_1(t_1)h_1(t_2)h_3^*(t_3)-2h_1(t_1)h_2(t_2)h_1^*(t_3)+4h_1(t_1)h_1^*(t_2)h_2(t_3)\\
&+4h_1(t_1)h_3^*(t_2)h_1(t_3)-2h_2(t_1)h_1(t_2)h_1^*(t_3)+4h_2(t_1)h_1^*(t_2)h_1(t_3)\\
&+-2h_3^*(t_1)h_1(t_2)h_1(t_3)-2h_1^*(t_1)h_2(t_2)h_1(t_3)-2h_1^*(t_1)h_1(t_2)h_2(t_3)\Big)\sigma_+a\\
&+\Big(-2h_1(t_1)h_1(t_2)h_2^*(t_3)-2h_1(t_1)h_3(t_2)h_1^*(t_3)+4h_1(t_1)h_1^*(t_2)h_3(t_3)\\
&+ 4h_1(t_1)h_2^*(t_2)h_1(t_3)-2h_3(t_1)h_1(t_2)h_1^*(t_3)+4h_3(t_1)h_1^*(t_2)h_1(t_3)\\
&+-2h_2^*(t_1)h_1(t_2)h_1(t_3)-2h_1^*(t_1)h_3(t_2)h_1(t_3)-2h_1^*(t_1)h_1(t_2)h_3(t_3)\Big)\sigma_+a^{\dagger}\\
&+h.c.+O(\eta^2,\eta^3)
\end{split}
\end{equation}
By integrating the above nested commutator, we will arrive the short hand expression for the second order effective Hamiltonian (Eq. \eqref{eq:shorthand_eff_2}. As explained in Appendix \ref{appendix:Effective Hamiltonian}, we have chosen to neglect terms such as $h_2^*(t_1)h_3(t_2)h_1(t_3)$ and $h_3^*(t_1)h_3(t_2)h_3(t_3)$ to avoid further complication. The remaining task is to evaluate the triple-integrals of all the terms in Eq. \eqref{eq:nested commutator}, and we refer interested readers to read Appendix \ref{appendix:List of Integrals} for the technical details.

\newpage
\section{Target Functional}\label{appendix:Target Functional}
To complete the discussion in subsection \ref{Target Functional}, we provide here the analytical expressions of the integrals in Eq. \eqref{eq:gate_infidelity_simplified} and Eq. \eqref{eq:state_infidelity_simplified}.
\begin{equation}
\frac{1}{T}\int_0^Tdt\, \,|g_1|^2=\sum_{j=-n}^{n}\frac{(\text{f}_j)^2}{4(m-j)^2}+\Bigg(\sum_{j=-n}^{n}\frac{\text{f}_j}{2(m-j)}\Bigg)^2\hspace{5cm}
\end{equation}
\begin{equation}
\frac{1}{T}\int_0^Tdt\, \,|g_2|^2=\sum_{\substack{j=-n\\(j\neq0)}}^{n}\frac{\eta_j^2 \,(\text{f}_j)^2}{4j^2}+\Bigg(\sum_{\substack{j=-n\\(j\neq 0)}}^{n}\frac{\eta_j\text{f}_j}{2j}\Bigg)^2\hspace{6cm}
\end{equation}
\begin{equation}
\frac{1}{T}\int_0^Tdt\, \,|g_3|^2=\sum_{j=-n}^{n}\frac{\eta_j^2 \,(\text{f}_j)^2}{4(2m-j)^2}+\Bigg(\sum_{j=-n}^{n}\frac{\eta_j\text{f}_j}{2(2m-j)}\Bigg)^2\hspace{4.5cm}
\end{equation}
\begin{equation}
\begin{split}
&\frac{1}{T}\int_0^Tdt\, \,\text{cos}(\text{f}_{tg}\sqrt{k}\omega t)|g_3|^2=\sum_{j,q=-n}^{n} \frac{i\eta_j \eta_q \text{f}_j \text{f}_q}{16\pi (2m-j)(2m-q)} \times\\
&\Bigg(  \frac{e^{2\pi i(2m-j+\text{f}_{tg}\sqrt{k})}-1}{(2m-j+\text{f}_{tg}\sqrt{k})}-\frac{e^{-2\pi i(2m-q-\text{f}_{tg}\sqrt{k})}-1}{(2m-q-\text{f}_{tg}\sqrt{k})}-\frac{e^{2\pi i\text{f}_{tg}\sqrt{k}}-1}{\text{f}_{tg}\sqrt{k}}-\frac{e^{2\pi i(q-j+\text{f}_{tg}\sqrt{k})}-1}{(q-j+\text{f}_{tg}\sqrt{k})}\Bigg)
\end{split}
\end{equation}
As suggested in section \ref{Numerical Results}, higher order corrections are needed for the investigation of the system under strong driving. In the following, we provide a more detailed discussion on how to expand the state infidelity (Eq. \eqref{eq:state_infidelity_simplified}) to higher order systematically. Due to periodicity, the propagator $\bar{U}$ is decomposed as $U_{F}U_{eff}$ according to the Floquet theory \cite{floquet1883equations}. The expression can be rearranged as $U_{F}=\bar{U}U_{eff}^{\dagger}$, and as a consequence, the fluctuation term $U_{F}$ can also be expanded as follow,
\begin{equation}
U_F=\text{exp}(-iG(t))=\text{exp}(-i[G^{(1)}+G^{(2)}+G^{(3)}+\cdots])
\end{equation}
where the expansion series can be found by applying the Baker-Cambell-Hausdorff formula (see the Supplemental Material for \cite{verdeny2014optimal}). The expansion of $G(t)$ provides a systematic way to expand the fluctuation term of the Floquet-decomposed propagator,
\begin{equation}
U_F=\mathds{1}-i(G^{(1)}+G^{(2)}+G^{(3)}+\cdots)-\frac{1}{2}(G^{(1)}+G^{(2)}+G^{(3)}+\cdots)^2+\cdots
\end{equation}
In subsection \ref{Target Functional}, the state infidelity is only approximated to second order, and we provide here an expansion of the functional up to fourth order.
\begin{equation}\label{eq:functional_higher_order}
\begin{split}
\mathcal{I}_{state}&\approx\frac{1}{T}\int^{T}_0 dt \, \Big(1- \braket{U_F}\braket{U_F^{\dagger}}\Big)\\
&=\frac{1}{T}\int^{T}_0 dt \, \Big(\braket{[G^{(1)}]^2}+\braket{\{G^{(1)},G^{(2)}\}}+\braket{[G^{(2)}]^2}+\braket{\{G^{(1)},G^{(3)}\}}-(\braket{G^{(2)}})^2-\frac{1}{12}\braket{[G^{(1)}]^4}-\frac{1}{4}(\braket{[G^{(1)}]^2})^2 \Big)
\end{split}
\end{equation}
where $\braket{A}$ is introduced here as a short hand notation for $\braket{i|U_{tg}^{\dagger}AU_{tg}|i}$. $\braket{G^{(1)}}$ is equal to zero for the system considered in this paper \footnote{To be clear on our symbols, $\braket{G^{(1)}}=0$ does not imply that $\braket{[G^{(1)}]^2}=0$}, so any term associated with $\braket{G^{(1)}}$ is not shown. By dropping the third and fourth order terms in Eq. \eqref{eq:functional_higher_order}, Eq. \eqref{eq:state_infidelity_simplified} in the main text can be recovered. The exact evaluation of Eq. \eqref{eq:functional_higher_order} is beyond the scope of this paper, but it would be of great interest in the future to investigate if the optimal pulse can be further improved by including high order terms. 
\newpage

\section{Further List of Integrals}\label{appendix:List of Integrals}
Following the discussion in Appendix \ref{appendix:Second Order Effective Hamiltonian}, we provide here the list of integrals which are used to evaluate the second order effective Hamiltonian. In the following, we provide the analytical expressions of $\alpha_{j}^{(2)}$ where $m>n$.
\begin{equation}
\begin{split}
&\alpha_{1}^{(2)}=\frac{-1}{6T} \int_{0}^{T}dt_{1}\int_{0}^{t_1}dt_2 \int_{0}^{t_2}dt_3 \, \Big((\delta \omega)^2 h_1(t_3)+(\delta \omega)^2 h_1(t_1)-2(\delta \omega)^2 h_1(t_2) \Big)\hspace{4.95cm}\\
&\text{For}\,\, m>n:\\
&\hspace{0.605cm}=-\sum_{j=-n}^{n} \frac{\delta^2 \text{f}_j \omega}{2(m-j)^2}\\
\end{split}
\end{equation}
\begin{equation}
\begin{split}
&\alpha_{2}^{(2)}=\frac{-1}{6T} \int_{0}^{T}dt_{1}\int_{0}^{t_1}dt_2 \int_{0}^{t_2}dt_3 \, \Big((\delta \omega)^2 h_2(t_3)+(\delta \omega)^2 h_2(t_1)-2(\delta \omega)^2 h_2(t_2) \Big)\hspace{4.95cm}\\
&\text{For}\,\, m>n:\\
&\hspace{0.605cm}=-\sum_{\substack{j=-n \\ (j \neq 0)}}^{n} \frac{i\delta^2 \eta_j \text{f}_j \omega}{2j^2}\\
\end{split}
\end{equation}
\begin{equation}
\begin{split}
&\alpha_{3}^{(2)}=\frac{-1}{6T} \int_{0}^{T}dt_{1}\int_{0}^{t_1}dt_2 \int_{0}^{t_2}dt_3 \, \Big((\delta \omega)^2 h_3(t_3)+(\delta \omega)^2 h_3(t_1)-2(\delta \omega)^2 h_3(t_2) \Big)\hspace{4.95cm}\\
&\text{For}\,\, m>n:\\
&\hspace{0.605cm}=-\sum_{j=-n}^{n} \frac{i\delta^2 \eta_j \text{f}_j \omega}{2(2m-j)^2}\\
\end{split}
\end{equation}
\begin{equation}
\begin{split}
&\alpha_{4}^{(2)}=\frac{-1}{6T} \int_{0}^{T}dt_{1}\int_{0}^{t_1}dt_2 \int_{0}^{t_2}dt_3 \, \Big(\delta \omega h_1^*(t_1)h_1(t_3)+\delta \omega h_1(t_1) h_1^*(t_3)-\delta \omega h_1^*(t_1)h_1(t_2)-\delta \omega h_1(t_2)h_1^*(t_3)\Big)\hspace{1.67cm}\\
&\text{For}\,\, m>n:\\
&\hspace{0.605cm}=\Big(\sum_{j=-n}^{n}\frac{\delta (\text{f}_j)^2}{8(m-j)^2}+\sum_{j,q=-n}^{n}\frac{\delta \text{f}_j \text{f}_q}{8(m-j)(m-q)}\Big)\omega\\ 
\end{split}
\end{equation}
\begin{equation}
\begin{split}
&\alpha_{5}^{(2)}=\frac{-1}{6T} \int_{0}^{T}dt_{1}\int_{0}^{t_1}dt_2 \int_{0}^{t_2}dt_3 \, \Big(+\delta \omega h_1^*(t_1)h_2(t_3)+\delta \omega h_2(t_1) h_1^*(t_3)-\delta \omega h_1^*(t_1)h_2(t_2)-\delta \omega h_2(t_2)h_1^*(t_3)\hspace{1.45cm}\\
&\hspace{5.42cm}+\delta \omega h_1(t_1)h_3^*(t_3)+\delta \omega h_3^*(t_1)h_1(t_3)-\delta \omega h_1(t_1)h_3^*(t_2)-\delta \omega h_3^*(t_2)h_1(t_3)\\
&\hspace{5.42cm} +\delta \omega h_2(t_1)h_1^*(t_3)+\delta \omega h_1^*(t_1)h_2(t_3)-\delta \omega h_2(t_1)h_1^*(t_2)-\delta \omega h_1^*(t_2)h_2(t_3)\\
&\hspace{5.42cm}+\delta \omega h_3^*(t_1)h_1(t_3)+\delta \omega h_1(t_1)h_3^*(t_3)-\delta \omega h_3^*(t_1)h_1(t_2)-\delta \omega h_1(t_2)h_3^*(t_3)\Big)\\
&\text{For}\,\, m>2n:\\
&\hspace{0.605cm}=\Big(-\sum_{j=-n}^{n}\frac{i\delta \eta_0 \text{f}_0 \text{f}_j}{4(m-j)^2}-\sum_{\substack{j,q=-n\\(q \neq 0)}}^{n}\frac{i\delta \eta_q \text{f}_j\text{f}_q}{4q(m-j)}-\sum_{j,q=-n}^{n}\frac{i\delta \eta_j \text{f}_j \text{f}_q}{4(m-q)(2m-j)}\Big)  \omega\\
&\text{For}\,\, 2n\geq m>n:\\
&\hspace{0.605cm}=\Bigg(-\sum_{j=-n}^{n}\frac{i\delta\eta_0\text{f}_0\text{f}_j}{4(m-j)^2}-\sum_{\substack{j,q=-n\\(q \neq 0)}}^{n}\frac{i\delta\eta_q\text{f}_j\text{f}_q}{4q(m-j)}+\sum_{\substack{j,q=-n\\(q \neq 0)\\(m-j+q=0)}}^{n}\frac{i\delta\eta_q\text{f}_j\text{f}_q}{8q^2}+\sum_{\substack{j,q=-n\\(m-q+j=0)}}^{n}\frac{i\delta\eta_j\text{f}_j\text{f}_q}{8(m-q)^2}\\
&\hspace{0.605cm}-\sum_{j,q=-n}^{n}\frac{i\delta\eta_j\text{f}_j\text{f}_q}{4(m-q)(2m-j)}-\sum_{\substack{j,q=-n\\(m+j-q=0)}}^{n}\frac{i\delta\eta_q\text{f}_j\text{f}_q}{6(2m-q)^2}-\sum_{\substack{j,q=-n\\(m-j+q=0)}}^{n}\frac{i\delta\eta_j\text{f}_j\text{f}_q}{12(m-q)^2}\Bigg)\omega
\end{split}
\end{equation}
\begin{equation}
\begin{split}
&\alpha_{6}^{(2)}=\frac{-1}{6T} \int_{0}^{T}dt_{1}\int_{0}^{t_1}dt_2 \int_{0}^{t_2}dt_3 \, \Big(+\delta \omega h_2(t_1)h_3^*(t_3)+\delta \omega h_3^*(t_1)h_2(t_3)-\delta \omega h_2(t_1)h_3^*(t_2)-\delta \omega h_3^*(t_2)h_2(t_3)\hspace{1.45cm}\\
&\hspace{5.42cm}+\delta \omega h_3^*(t_1)h_2(t_3)+\delta \omega h_2(t_1)h_3^*(t_3)-\delta \omega h_3^*(t_1)h_2(t_2)-\delta \omega h_2(t_2)h_3^*(t_3)\Big)\\
&\text{For}\,\, m>n:\\
&\hspace{0.605cm}=\Big(-\sum_{j=-n}^{n} \frac{\delta \eta_0 \eta_j \text{f}_0 \text{f}_j}{4(2m-j)^2}-\sum_{\substack{j,q=-n\\(j \neq 0)}}^{n}\frac{\delta \eta_j \eta_q \text{f}_j\text{f}_q}{4j(2m-q)}\Big)\omega
\end{split}
\end{equation}
\begin{equation}
\begin{split}
&\alpha_{7}^{(2)}=\frac{-1}{6T} \int_{0}^{T}dt_{1}\int_{0}^{t_1}dt_2 \int_{0}^{t_2}dt_3 \, \Big(+\delta \omega h_2^*(t_1)h_2(t_3)+\delta \omega h_2(t_1)h_2^*(t_3)-\delta \omega h_2^*(t_1)h_2(t_2)-\delta \omega h_2(t_2)h_2^*(t_3)\Big)\hspace{1.45cm}\\
&\text{For}\,\, m>n:\\
&\hspace{0.605cm}=\Big(\sum_{\substack{j=-n\\(j \neq 0)}}^{n}\frac{\delta (\eta_j)^2 (\text{f}_j)^2}{8j^2}-\sum_{\substack{j=-n\\(j \neq 0)}}^{n}\frac{\delta \eta_0 \eta_j \text{f}_0\text{f}_j}{4j^2}+\sum_{\substack{j,q=-n\\(j,q \neq 0)}}^{n}\frac{\delta \eta_j \eta_q \text{f}_j \text{f}_q}{8jq}\Big)\omega
\end{split}
\end{equation}
\begin{equation}
\begin{split}
&\alpha_{8}^{(2)}=\frac{-1}{6T} \int_{0}^{T}dt_{1}\int_{0}^{t_1}dt_2 \int_{0}^{t_2}dt_3 \, \Big(+\delta \omega h_3^*(t_1)h_3(t_3)+\delta \omega h_3(t_1) h_3^*(t_3)-\delta \omega h_3^*(t_1)h_3(t_2)-\delta \omega h_3(t_2)h_3^*(t_3)\Big)\hspace{1.45cm}\\
&\text{For}\,\, m>n:\\
&\hspace{0.605cm}=\Big(\sum_{j=-n}^{n}\frac{\delta(\eta_j)^2(\text{f}_j)^2}{8(2m-j)^2}+\sum_{j,q=-n}^{n}\frac{\delta\eta_j \eta_q \text{f}_j \text{f}_q}{8(2m-j)(2m-q)}\Big)\omega
\end{split}
\end{equation}
\begin{equation}\label{h,eff,9,(2)}
\begin{split}
&\alpha_{9}^{(2)}=\frac{-1}{6T} \int_{0}^{T}dt_{1}\int_{0}^{t_1}dt_2 \int_{0}^{t_2}dt_3 \, \Big( -2h_1(t_1)h_1(t_2)h_1^*(t_3)+4h_1(t_1)h_1^*(t_2)h_1(t_3)-2h_1^*(t_1)h_1(t_2)h_1(t_3)\Big)\hspace{2.25cm}\\
&\text{For}\,\, m>3n:\\
&\hspace{0.605cm}=\Big(\sum_{j,q=-n}^{n}\frac{\text{f}_j(\text{f}_q)^2}{4(m-j)(m-q)}\Big)\omega\\
&\text{For}\,\, 3n\geq m>n:\\
&\hspace{0.605cm}=\Big(\sum_{j,q=-n}^{n}\frac{\text{f}_j(\text{f}_q)^2}{4(m-j)(m-q)}+\sum_{\substack{j,q,r=-n\\(q \neq r)\\(m-j+q-r=0)}}^{n}\frac{\text{f}_j \text{f}_q\text{f}_r}{8(m-r)(q-r)}\Big)\omega\\
\end{split}
\end{equation}
\begin{equation}\label{h,eff,10,(2)}
\begin{split}
&\alpha_{10}^{(2)}=\frac{-1}{6T} \int_{0}^{T}dt_{1}\int_{0}^{t_1}dt_2 \int_{0}^{t_2}dt_3 \, \Big( -2h_1(t_1)h_1(t_2)h_3^*(t_3)-2h_1(t_1)h_2(t_2)h_1^*(t_3)+4h_1(t_1)h_1^*(t_2)h_2(t_3)\hspace{2.25cm}\\
&\hspace{5.42cm}+4h_1(t_1)h_3^*(t_2)h_1(t_3)-2h_2(t_1)h_1(t_2)h_1^*(t_3)+4h_2(t_1)h_1^*(t_2)h_1(t_3)\\
&\hspace{5.42cm}-2h_3^*(t_1)h_1(t_2)h_1(t_3)-2h_1^*(t_1)h_2(t_2)h_1(t_3)-2h_1^*(t_1)h_1(t_2)h_2(t_3)\Big)\\
&\text{For}\,\, m>2n:\\
&\hspace{0.605cm}=\Big(-\sum_{\substack{j,q,r=-n\\(q-r-j=0)}}^{n} \frac{i\eta_q\text{f}_j \text{f}_q \text{f}_r \omega}{8(m-r)(q-r-m)}-\sum_{\substack{j,q,r=-n\\(q-r+j=0)}}^{n} \frac{i\eta_q\text{f}_j \text{f}_q \text{f}_r \omega}{12(m-r)(m+q-r)}-\sum_{j,q=-n}^{n}\frac{i\eta_0 \text{f}_0 \text{f}_j \text{f}_q \omega}{6(m-j)(m-q)}\\
&\hspace{0.605cm}-\sum_{\substack{j,q=-n\\(j\neq0)}}^{n} \frac{i\eta_j \text{f}_j (\text{f}_q)^2\omega}{4j(m-q)}-\sum_{\substack{j,q,r=-n\\(q\neq r)\\(r+j-q=0)}}^{n} \frac{i\eta_j \text{f}_j \text{f}_q \text{f}_r\omega}{6(m-r)(r-q) }+\sum_{\substack{j,q=-n\\(j\neq q)}}^{n} \frac{i\eta_0 \text{f}_0 \text{f}_j \text{f}_q \omega}{12(m-q)(q-j)}+\sum_{\substack{j,q,r=-n\\(q\neq r)\\(r-q-j=0)}}^{n} \frac{i\eta_j \text{f}_j \text{f}_q \text{f}_r \omega}{12(m-r)(r-q)}\Big)\omega\\
&\text{For}\,\,2n\geq m>n:\\
&\hspace{0.605cm}=\Big(-\sum_{\substack{j,q,r=-n\\(j-m-q \neq 0)\\(j-r-q=0)}}^{n}\frac{i \eta_j \text{f}_j\text{f}_q\text{f}_r}{4(2m-j)(j-m-q)}-\sum_{\substack{j,q,r=-n\\(j-m-q=0)}}^{n}\frac{i \eta_j \text{f}_j\text{f}_q\text{f}_r}{4(2m-j)(m-r)}+\sum_{\substack{j,q,r=-n\\(q\neq r)\\(r-q-j=0)}}^{n}\frac{i\eta_j\text{f}_j\text{f}_q\text{f}_r}{12(m-r)(r-q)}\\
&\hspace{0.605cm}-\sum_{\substack{j,q,r=-n\\(m-r+j \neq 0)\\(r-j-q=0)}}^{n}\frac{i \eta_j \text{f}_j\text{f}_q\text{f}_r}{12(m-r)(m-r+j)}+\sum_{\substack{j,q,r=-n\\(m-r+j=0)}}^{n}\frac{i \eta_j \text{f}_j\text{f}_q\text{f}_r}{12(m-r)(m-q)}+\sum_{\substack{j,q,r=-n\\(j \neq 0)\\(m-j-q=0)}}^{n}\frac{i \eta_j \text{f}_j\text{f}_q\text{f}_r}{12j(m-r)}\\
&\hspace{0.605cm}-\sum_{\substack{j,q=-n\\(j\neq q)}}^{n}\frac{i\eta_0\text{f}_0\text{f}_j\text{f}_q}{4(m-j)(j-q)}-\sum_{\substack{j,q=-n\\(j \neq 0)}}^{n}\frac{i \eta_j \text{f}_j (\text{f}_q)^2}{4j(m-q)}-\sum_{\substack{j,q,r=-n\\(m-q-j=0)}}^{n}\frac{i\eta_j\text{f}_j\text{f}_q\text{f}_r}{12(m-r)(m-q)}\\
&\hspace{0.605cm}-\sum_{j=-n}^{n}\frac{i\eta_0\text{f}_0(\text{f}_j)^2}{6(m-j)^2}-\sum_{\substack{j,q,r=-n\\(j \neq 0)\\(m-q+j=0)}}^{n}\frac{i \eta_j \text{f}_j\text{f}_q\text{f}_r}{6j(m-r)}+\sum_{\substack{j,q,r=-n\\(j \neq 0)\\(m-q+j\neq0)\\(q-r-j=0)}}^{n}\frac{i \eta_j \text{f}_j\text{f}_q\text{f}_r}{6j(m-q+j)}\Big)\omega\\
\end{split}
\end{equation}
\begin{equation}\label{h,eff,11,(2)}
\begin{split}
&\alpha_{11}^{(2)}=\frac{-1}{6T} \int_{0}^{T}dt_{1}\int_{0}^{t_1}dt_2 \int_{0}^{t_2}dt_3 \, \Big( -2h_1(t_1)h_1(t_2)h_2^*(t_3)-2h_1(t_1)h_3(t_2)h_1^*(t_3)+4h_1(t_1)h_1^*(t_2)h_3(t_3)\hspace{2.25cm}\\
&\hspace{5.42cm}+ 4h_1(t_1)h_2^*(t_2)h_1(t_3)-2h_3(t_1)h_1(t_2)h_1^*(t_3)+4h_3(t_1)h_1^*(t_2)h_1(t_3)\\
&\hspace{5.42cm}-2h_2^*(t_1)h_1(t_2)h_1(t_3)-2h_1^*(t_1)h_3(t_2)h_1(t_3)-2h_1^*(t_1)h_1(t_2)h_3(t_3)\Big)\\
&\text{For}\,\, m>2n:\\
&\hspace{0.605cm}=\Big(\sum_{j,q=-n}^{n} \frac{i\eta_0 \text{f}_0 \text{f}_j \text{f}_q \omega}{8(m-j)(m-q)}+\sum_{j,q=-n}^{n} \frac{i\eta_j \text{f}_j (\text{f}_q)^2 \omega}{4(m-q)(2m-j)}\Big)\omega\\
&\text{For}\,\, 2n\geq m>\frac{3}{2}n:\\
&\hspace{0.605cm}=\Big(\sum_{j,q=-n}^{n}\frac{i\eta_0\text{f}_0\text{f}_j\text{f}_q}{4(m-q)(2m-j-q)}+\sum_{\substack{j,q,r=-n\\(j \neq 0)\\(m-q+j=0)}}^{n}\frac{i \eta_j \text{f}_j\text{f}_q\text{f}_r}{4j(m-r)}+\sum_{j,q=-n}^{n}\frac{i\eta_j\text{f}_j(\text{f}_q)^2}{4(m-q)(2m-j)}\\
&\hspace{0.605cm}+\sum_{\substack{j,q,r=-n\\(m+q-j=0)}}^{n}\frac{i \eta_j \text{f}_j\text{f}_q\text{f}_r}{4(2m-j)(m-r)}\Big)\omega\\
&\text{For}\,\, \frac{3}{2}n>m>n:\\
&\hspace{0.605cm}=\Big(\sum_{j,q=-n}^{n}\frac{i\eta_0\text{f}_0\text{f}_j\text{f}_q}{4(m-q)(2m-j-q)}+\sum_{\substack{j,q,r=-n\\(j \neq 0)\\(m-q+j=0)}}^{n}\frac{i \eta_j \text{f}_j\text{f}_q\text{f}_r}{4j(m-r)}+\sum_{\substack{j,q,r=-n\\(j \neq 0)\\(m-q+j\neq0)\\(2m-r-q+j=0)}}^{n}\frac{i \eta_j \text{f}_j\text{f}_q\text{f}_r}{4j(m-q+j)}\\
&\hspace{0.605cm}+\sum_{j,q=-n}^{n}\frac{i\eta_j\text{f}_j(\text{f}_q)^2}{4(m-q)(2m-j)}+\sum_{\substack{j,q,r=-n\\(q \neq r)\\(2m-j+r-q=0)}}^{n}\frac{i \eta_j \text{f}_j\text{f}_q\text{f}_r}{12(m-r)(r-q)}+\sum_{\substack{j,q,r=-n\\(m+q-j=0)}}^{n}\frac{i \eta_j \text{f}_j\text{f}_q\text{f}_r}{4(2m-j)(m-r)}\\
&\hspace{0.605cm}+\sum_{\substack{j,q,r=-n\\(m+q-j\neq0)\\(2m-j+q-r=0)}}^{n}\frac{i \eta_j \text{f}_j\text{f}_q\text{f}_r}{6(2m-j)(m+q-j)}-\sum_{\substack{j,q,r=-n\\(2m+q-j-r=0)}}^{n}\frac{i \eta_j \text{f}_j\text{f}_q\text{f}_r}{12(m-r)(3m-j-r)}\Big)\omega
\end{split}
\end{equation}


\twocolumngrid



\end{document}